\documentclass[twocolumn,amsmath,amssymb,aps,]{revtex4-1}
\usepackage{bm}
\usepackage[colorlinks=true,urlcolor=blue,linkcolor=blue,citecolor=blue]{hyperref}
\usepackage{color}
\usepackage{graphics}
\usepackage{graphicx}
\usepackage{epsfig}
\usepackage{amssymb}
\usepackage{amsmath}
\usepackage{hyperref}
\usepackage{physics}

\usepackage{array}                 %Más entornos para tablas y arrays: https://www.ctan.org/pkg/array

\usepackage{amsfonts}           %Fuentes y símbolos: https://www.ctan.org/pkg/amsfonts

\begin{document}

\preprint{AIP/123-QED}

\title[Cyclotron-Frequency Stability via Single-Ion Fluorescence in a Penning Trap with a Cryogen-Free Superconducting Magnet]{Cyclotron-Frequency Stability via Single-Ion Fluorescence in a Penning Trap with a Cryogen-Free Superconducting Magnet }
 
% Force line breaks with \\
\author{A. Schupeta$^1$}
\author{D. Yousaf$^{1,2}$}
\author{M. Almagro$^1$}
\author{M. Hurtado$^{3,4}$}
\author{J.~M. Palomino$^4$}
\author{J. Berrocal$^1$}\thanks{Present address: National Institute of Standards and Technology (NIST), Boulder, United States}
\author{M. Block$^{5,6,7}$}
\author{Ch. E. D\"ullmann$^{5,6,7}$}
\author{D.~Rodr\'iguez$^{1,2,8}$}\email[Corresponding author: ]{danielrodriguez@ugr.es}

\affiliation{ 
$^1$Departamento de Física Atómica, Molecular y Nuclear, Universidad de Granada, 18071 Granada, Spain\\
$^2$Laboratorio de Trampas de Iones y L\'aseres, Universidad de Granada, 18071 Granada, Spain\\
$^3$Ultrasonics Laboratory, Universidad de Granada, 18071 Granada, Spain\\
$^4$Departamento de Mec\'anica de Estructuras e Ingenier\'ia Hidr\'aulica, Universidad de Granada, 18071 Granada, Spain\\
$^5$Department Chemie, Johannes Gutenberg-Universit\"at Mainz, 55099 Mainz, Germany\\
$^6$Helmholtz-Institut Mainz, 55099,  Mainz, Germany\\
$^7$GSI Helmholtzzentrum f\"ur Schwerionenforschung GmbH, 64291, Darmstadt, Germany\\
$^8$Centro de Investigación en Tecnologías de la Información y las Comunicaciones, Universidad de Granada, 18071 Granada, Spain
}

\date{\today}% It is always \today, today,

\begin{abstract}
In this work, we investigate the magnetic-field stability of a Penning trap operated with a cryogen-free superconducting magnet through cyclotron-frequency measurements of a single laser-cooled calcium ion using two complementary photon-based detection techniques: a pulsed optical method for determining the three ion eigenfrequencies and Fluorescence-Detected Fourier-Transform Ion-Cyclotron-Resonance (FD-FT-ICR). For the former technique, the data-analysis procedure was revisited, considering it as the dominant contribution to the measurement uncertainty and yielding a long-term magnetic-field drift $(1/B)(dB/dt) = -3.07(32)\times10^{-9}$~$\mathrm{h}^{-1}$. This value is comparable to those reported for high-precision Penning-trap experiments employing liquid-helium-based superconducting magnets. The short-term stability was investigated using the FD-FT-ICR technique, resulting in a minimum relative magnetic-field variation of $\delta B/B \simeq 5\times10^{-8}$ for averaging times between 30 and 60~s. Furthermore, this technique enables direct cyclotron-frequency determinations on timescales as short as a few seconds, providing access to magnetic-field fluctuations that are generally not resolved in conventional Penning-trap experiments.
\end{abstract}

\maketitle

\begin{center}
\small
The following article has been submitted to Review of Scientific
Instruments. %After it is published, it will be found at
%https://pubs.aip.org/rsi.
\end{center}

\section{\label{sec:level1} Introduction}

Cryogen-free Penning traps provide an alternative to conventional liquid-helium-based systems used in precision ion-trapping experiments, where superconducting magnets are typically operated in liquid-helium cryostats (see, e.g., \cite{Hann2008,Myer2015,Ulme2015,Gilm2017,Ball2019,Amsl2021,Schu2020,Corn2021}; see also Ref.~\cite{Blau2013} for a review of Penning-trap precision mass spectrometry at radioactive-ion-beam facilities). In mass spectrometry, the performance of cryogen-free Penning traps is particularly relevant for applications requiring high sensitivity and precision, which in some cases can only be achieved through appropriate single-ion detection techniques \cite{Hann2008,Myer2015,Ulme2015,Schu2020}. In laser-based experiments, however, single trapped ions can be detected via fluorescence imaging only for certain ion species. Within this context, a recent Penning-trap mass-spectrometry experiment employing laser-based detection has been demonstrated in Ref.~\cite{Berr2024}. This combination offers promising prospects for studies of superheavy elements (SHEs), which are produced in extremely small quantities. High-precision mass spectrometry and laser spectroscopy provide complementary information on nuclear binding energies, nuclear moments, and atomic structure \cite{Bloc2010,Mina2012,Kale2022,Laat2016,Raed2018,Lant2024}, important for addressing open questions highlighted in recent reviews (see, e.g., \cite{Giul2019,Smit2024}): how far the periodic table predicted by quantum mechanics may extend, how relativistic effects influence atomic structure, and the location of the so-called island of enhanced stability around $Z\sim114$ and $N\sim184$. 

At present, the limited efficiency of current experimental systems makes single-ion detection and manipulation essential for future experiments, requiring the development of a dedicated platform, as for example the one at the University of Granada (UGR) \cite{Guti2019,Berr2021,Berr2025}, which is the subject of this paper. Current experiments at UGR aim to form a $^{232}$Th$^+$ - $^{40}$Ca$^+$ hybrid crystal and to study sympathetic cooling via the laser-cooled $^{40}$Ca$^+$ ion, with the goal of performing purely photon-based mass spectrometry \cite{Guti2019b}, potentially reaching the quantum regime \cite{Cerr2021}. Thorium ions are of particular interest due to their connection to the low-energy nuclear isomer $^{229\hbox{\scriptsize{m}}}$Th, whose transition energy lies in the vacuum-ultraviolet range and has been proposed as the basis for a nuclear clock \cite{Peik2021,Thir2024}. To date, experiments on single thorium ions have primarily been carried out in Paul traps \cite{Camp2011,Groo2019,Zitz2025}.

$^{229}$Th is also of interest in its molecular form $^{229}$ThO$^+$, together with other diatomic molecules, as a candidate system to probe electroweak interactions through measurements of parity violation (PV) in a cryogen-free Penning trap \cite{Kart2024}. In such schemes, the magnetic field can be tuned to bring opposite-parity rotational and hyperfine states close to degeneracy via Zeeman shifts, thereby enhancing the weak-interaction-induced mixing and increasing the sensitivity to electroweak nuclear properties. These experiments require sympathetic laser cooling and single-ion detection.

In this work, we report on the characterization of cyclotron-frequency stability in a cryogen-free Penning trap using purely photon-based detection techniques on a laser-cooled Ca$^+$ ion. The results quantify the performance of the upgraded experimental platform and constitute an important step toward future precision experiments on heavy atomic and molecular ions. The setup at UGR has previously relied on a liquid-helium-based magnet to provide a highly homogeneous 7-T magnetic field. 

\section{Penning trap eigenfrequencies}
In an ideal hyperbolic Penning trap, a strong homogeneous magnetic field along the axial direction,~$\vec{B}=B\hat{e}_z$, provides confinement in the radial plane \cite{Brow1986}. By applying a potential difference between the endcaps and ring electrodes, and defining the characteristic trap dimension $d_0$, the electric potential $\Phi=(V_0/d_0^2)\left(2z^2-x^2-y^2\right)$, which provides confinement along the axial direction, has a quadrupolar form and cylindrical symmetry around the $z$-axis. A similar configuration can be realized using a stack of cylindrical electrodes \cite{Gabr1989}.

Due to Lorentz force, a charged particle with mass $m$ and charge state $q$ in the Penning trap undergoes a motion that can be decomposed into three eigenmotions: one along the axial direction, with frequency
\begin{equation}
\nu_z= \frac{1}{2\pi}\sqrt{\frac{qV_0}{md_0^2}},
\label{eq:axial_frequency}
\end{equation}
and two in the radial plane, with frequencies
\begin{equation}
\nu_\pm = \frac{\nu_c}{2}\left(1 \pm \sqrt{1-2\left(\frac{\nu_z}{\nu_c}\right)^2}\right),
\label{eq:redcyc_mag}
\end{equation}
where the subscripts $+$ and $-$ stand for modified-cyclotron and magnetron motion, respectively, and \linebreak $\nu_c =(1/2\pi)\cdot qB/m$ is the cyclotron frequency of the particle in the bare magnetic field. It is related to the radial eigenfrequencies through
\begin{equation}
\nu_c = \nu_+ + \nu_-.
\label{eq:cyclotron_direct}
\end{equation}
This relation is valid only up to a certain level of precision. However, the invariance theorem~\cite{Brow1982} given by
\begin{equation}
\nu_c^2 = \nu_+^2+\nu_-^2+\nu_z^2,
\label{eq:invariance_theorem}
\end{equation}
remains valid even in the presence of first-order misalignment between the magnetic and electric fields, as well as field ellipticity.

The dynamics of two ions simultaneously trapped in a Penning trap includes the Coulomb interaction.  In the case of two identical ions, %The ions are labeled as sensor with the subscript $s$ and target with the subscript $t$. We define $\mu=m_t/m_s$ and $\kappa =q_\mathrm{t}/q_\mathrm{s}$. 
when the kinetic energy is sufficiently low, e.g., after Doppler cooling, and $\nu_{z}/\nu_{c}<1/\sqrt{3}$, they form a Coulomb crystal aligned along the magnetic-field direction, with an equilibrium separation \cite{Mori2001,Guti2019}
\begin{equation}
d_\textrm{ion-ion} = \left(\frac{q^2}{8\pi^2\epsilon_0m_{s}\nu_{z}^2}\right)^{1/3},
\label{eq:crystal_equilibrium_distance}
\end{equation}
information that is used to obtain the magnification of the optical system, which is important for the photon-based techniques employed in this paper \cite{Berr2024,Berr2025}. 

\section{Experimental setup: Magnet, laser system and Penning trap}\label{sec:setup}

\begin{figure}[t]
\includegraphics[scale=0.40]{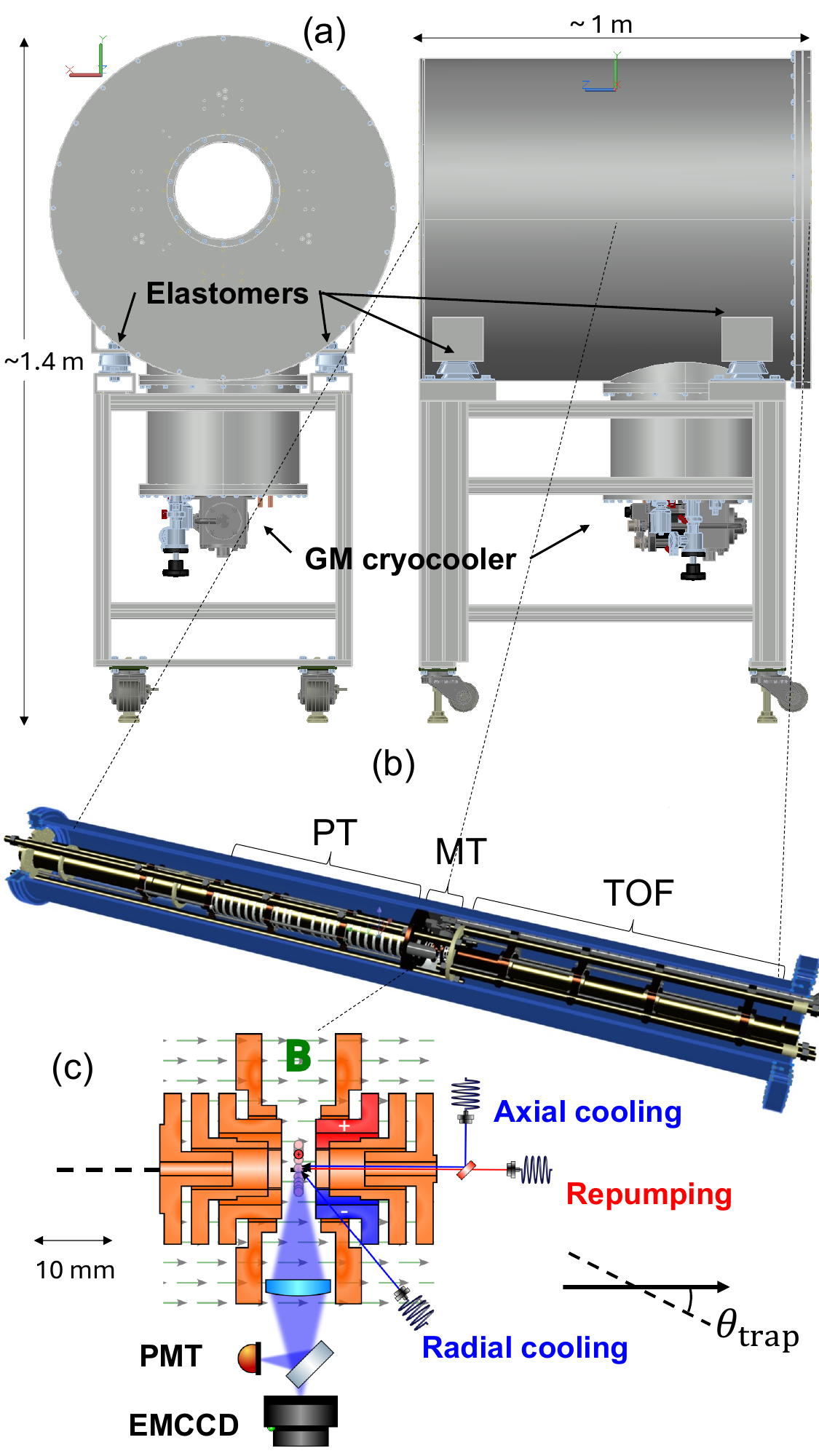}% Here is how to import EPS art
\vspace{-3mm}\caption{a) 3D-CAD drawings showing two views of the cryogen-free superconducting magnet mounted on its supporting structure. The cryocooler and elastomers on the ITEM structure are indicated. The magnet bore has a diameter of 240 mm. b) Vacuum tube housing the Penning trap system within the magnet bore. The tube is not attached to the magnet itself but is supported by custom-designed optical tables. c) Enlarged schematics view of the open-ring trap positioned at the center of the magnet bore. The first optical element of the imaging system objective, fluorescence detectors, and laser beams directions are also indicated. The dashed line is the symmetry axis of the trap. The angle $\theta _{\hbox{\scriptsize{trap}}}$ shown in the lower right corner accounts for the deviation of this axis from the magnetic field axis (black arrow). \label{fig:setup}}
\end{figure}

A three-dimensional CAD model of the magnet and Penning-trap system is shown in Figs.~\ref{fig:setup}(a) and (b). The \linebreak 7-T cryogen-free superconducting magnet houses two Penning traps: a Preparation Trap (PT)~\cite{Corn2016c}, which was not used in the present measurements, and the Measurement Trap (MT)~\cite{Guti2019}. A zoomed view of the MT and some elements is shown in Fig.~\ref{fig:setup}(c). Upstream of the magnet, outside the field of view of the figure, a Paul trap and a laser-ablation ion source are installed~\cite{Berr2022}. Ions generated in these devices are transported to the Penning traps through a dedicated transfer beamline. On the downstream side, only partly shown in the figure, a time-of-flight (TOF) section equipped with a movable microchannel-plate detector is used for destructive ion detection and TOF-based identification. The optical components for laser-beam preparation and the photon-detection system are located outside the vacuum chamber, beyond the TOF section.

The cryogen-free superconducting magnet from Scientific Magnetics (7T240) was energized in February 2023 for the first time. It features one region with a field homogeneity of 0.1~ppm in a $15\times15\times15$~mm$^{3}$ cube and 2~ppm in a $15\times15\times60$~mm$^{3}$ parallelepiped, both centered in the magnet's bore. A support assembly based on two U-shaped customized optical tables placed at both sides of the magnet has been implemented to hold the whole beamline. Such a system includes a rail configuration that facilitates the maintenance operations in the whole Penning-trap beamline and minimizes the relative misalignment between the beam-access optics~(mounted before on an independent optical breadboard) and the MT. 

 The magnet is cooled using a Gifford–McMahon (GM) cryocooler (model SRP-82 from SUMITOMO) indicated in Fig.~\ref{fig:setup}(a). This introduces unwanted mechanical vibrations into the system. Although the tube housing the traps (Fig.~\ref{fig:setup}(b)) is mechanically decoupled from the magnet bore, vibrations of the magnet can still influence the homogeneity of the magnetic field. The vibrational characteristics of the system, including the frequency spectrum and displacement amplitudes, were investigated with the cryocooler both operating and switched off using two complementary techniques: accelerometry based on MEMS (Micro-Electro-Mechanical Systems) sensors and laser Doppler vibrometry. The former method involved mounting a triaxial accelerometer directly on the magnet to obtain the vibration spectrum, whereas the latter was used to measure vibration amplitudes, with the laser beam oriented along both the horizontal and vertical directions.

The maximum displacements measured by laser Doppler vibrometry with the cryocooler operating were 1.87~$\mu$m and 3.64~$\mu$m in the horizontal and vertical directions, respectively, below the specifications provided by the manufacturer of the magnet. When the cryocooler was switched off, these values decreased to 0.39 and 0.31~$\mu$m. The fast Fourier transform (FFT) spectrum obtained from the accelerometer measurements exhibits a fundamental frequency at 1 Hz together with its harmonics, which are attributed to the compression–expansion cycle of the cryocooler. In addition, the FFT spectra derived from the laser-Doppler-vibrometry measurements reveal pronounced peaks at 20 and 99~Hz when the cryocooler is operating, which are absent when it is switched off.

\begin{figure}[b]
\includegraphics[scale=0.45]{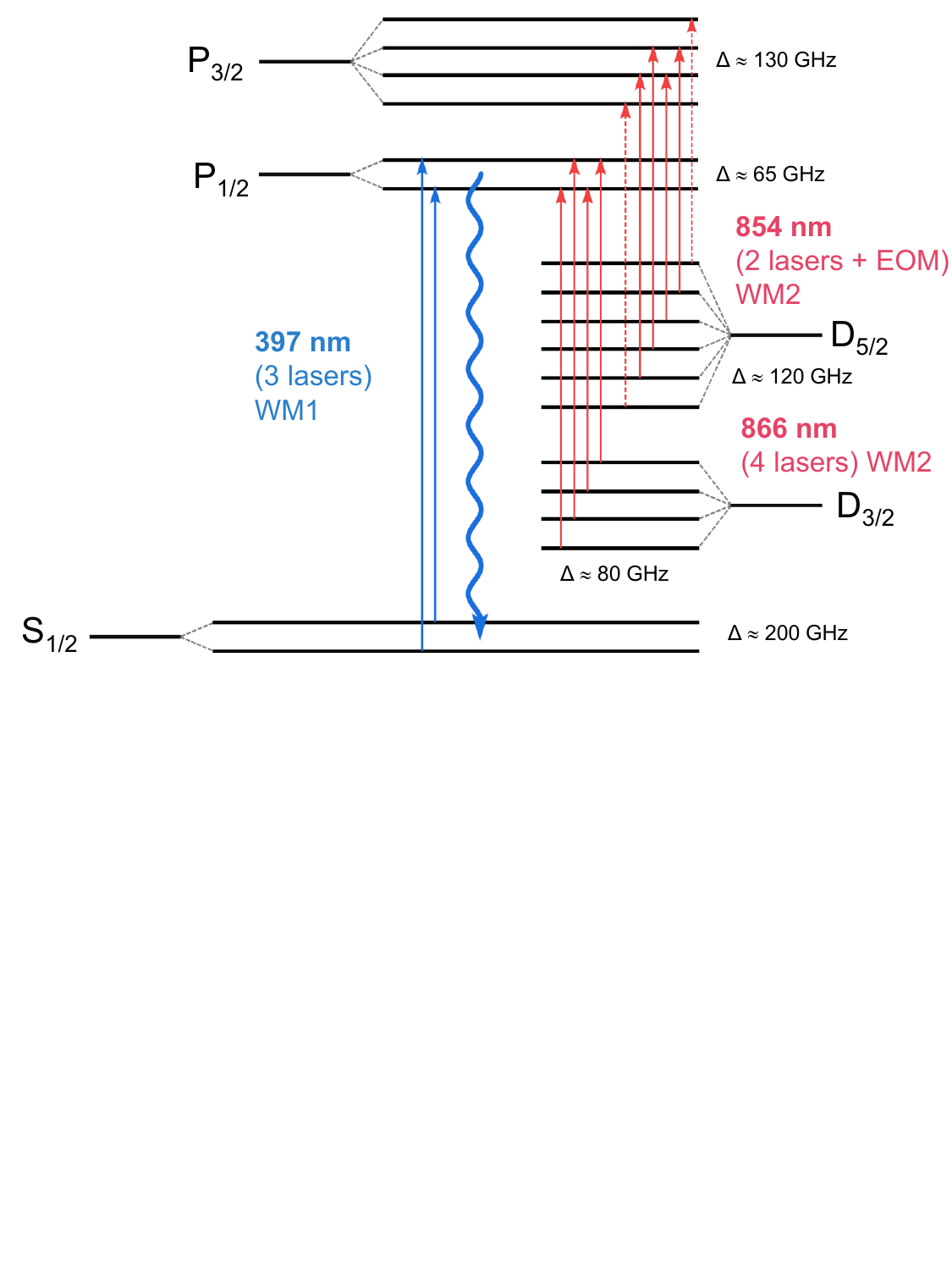}% Here is how to import EPS art
\vspace{-54mm}\caption{Relevant transitions of the even calcium isotopes in a 7-T magnetic field. Both the first-order (symmetric) and second-order (asymmetric) contributions to the Zeeman splitting are taken into account, with their approximate magnitudes indicated by $\Delta $. WM1 and WM2 denote Wave Meter 1 and Wave Meter 2, respectively. Dotted arrows represent transitions that were not required in the experiment, although laser light at these frequencies was provided in some configurations. \label{fig:level_scheme}}
\end{figure}

%\subsection{The open-ring Penning trap and the optical system}

The laser light required to drive the Doppler-cooling transitions shown in Fig.~\ref{fig:level_scheme} is generated by nine tunable external-cavity diode lasers. %(ECDLs). 
At least ten optical transitions must be addressed to efficiently cool the calcium ions to the Doppler limit \cite{Guti2019,Berr2022}. The system employs three 397-nm lasers to deliver two beams for addressing the radial ion motion and two for the axial motion, four 866-nm repumping lasers, and another two 854-nm repumping lasers (due to \linebreak $J$-mixing \cite{Cric2010}) which are phase-modulated using an electro-optical modulator (EOM). All laser frequencies are actively stabilized with an absolute accuracy better than 10~MHz by means of two  %Approximately 10\% of the output power from each laser is directed to 
wavelength meters (WLMs) %through optical switches, enabling sequential monitoring of multiple laser frequencies. The WLMs are 
that are referenced to a stabilized helium-neon laser. Frequency stabilization is achieved using a proportional-integral-derivative control loop, which calculates the correction voltage applied to the laser piezoelectric actuator through a digital-to-analog converter. 

The four 397-nm beams are delivered independently to the Penning-trap setup via single-mode optical fibers. The four 866-nm beams are combined into a single fiber, while the two 854-nm beams are coupled into a second fiber. Two of the 866-nm beams pass through an acousto-optic modulator operated in a single-pass configuration, allowing the cooling light to be switched on and off within 1~$\mu$s. % during the application of an external radio-frequency drive used to probe the ion eigenmotions. 
The axial and radial laser beams are prepared and combined on two dedicated optical tables. The radial beam contains only 397-nm light, whereas the axial beam carries the full set of wavelengths required for Doppler cooling and state repumping.

\begin{figure}[t]
\includegraphics[scale=0.25]{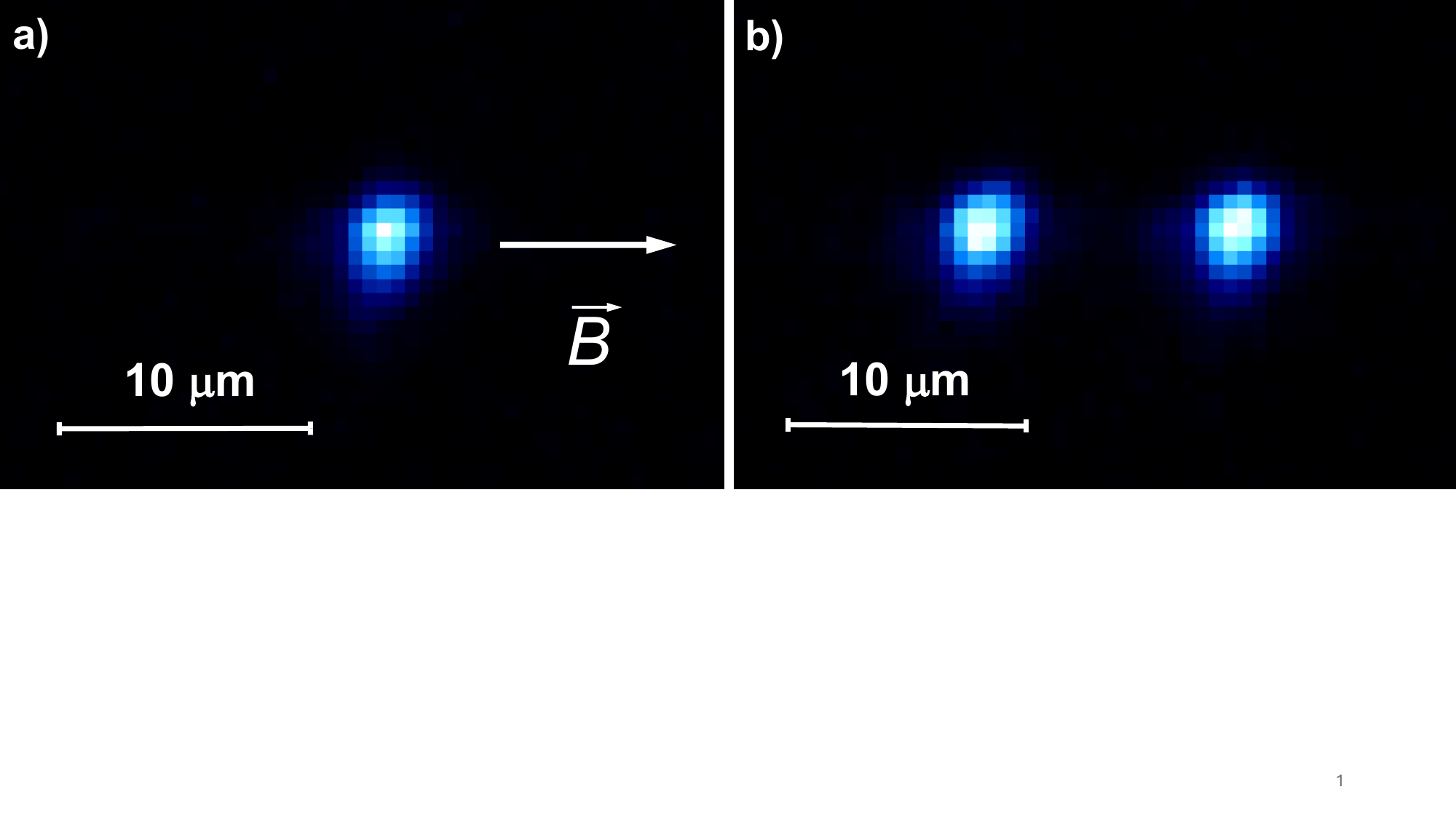}% Here is how to import EPS art
\vspace{-20mm}\caption{EMCCD images of a single $^{40}$Ca$^+$ ion (a) and a $^{40}$Ca$^+$-$^{40}$Ca$^+$ Coulomb crystal (b) for a trap depth of 30~V.\label{fig:excitation2}}
\end{figure}

The MT is an open-ring Penning trap (Fig.~\ref{fig:setup}(c)) described in detail in Ref.~\cite{Guti2019}. It is composed of two symmetric assemblies of four concentric electrodes, namely an endcap electrode, a ring electrode, a correction electrode, and a ground electrode. A schematic view of the trap is shown in Fig.~\ref{fig:setup}(c). The voltages applied to these electrodes can be adjusted to match the kinetic energy of the incoming ions. The ring electrode is segmented into four sectors, enabling the application of external RF fields in either dipolar (depicted in the figure) or quadrupolar configurations. The trap geometry provides both axial and radial optical access. Fluorescence photons emitted by the trapped ion are collected along the radial direction. The first element of the optical system is positioned 22~mm from the trap center, corresponding to a collection efficiency of 2.1~\%~\cite{Berr2025}. The optical system magnifies, and relays both the axial and radial projections from the trap center to the photon detectors; an Electron Multiplying Charge-Coupled Device (EMCCD) and a Photo Multiplier Tube (PMT), located outside the vacuum chamber. The fluorescence-photon signal is split in two in front of them. Figure~\ref{fig:excitation2} shows EMCCD images of a single ion (a) and a two-ion Coulomb crystal (b). The latter, together with Eq.~(\ref{eq:crystal_equilibrium_distance}), is used to determine the magnification of the imaging system, yielding a value of $\times 19.7$.

\begin{figure}[t]
\includegraphics[scale=0.26]{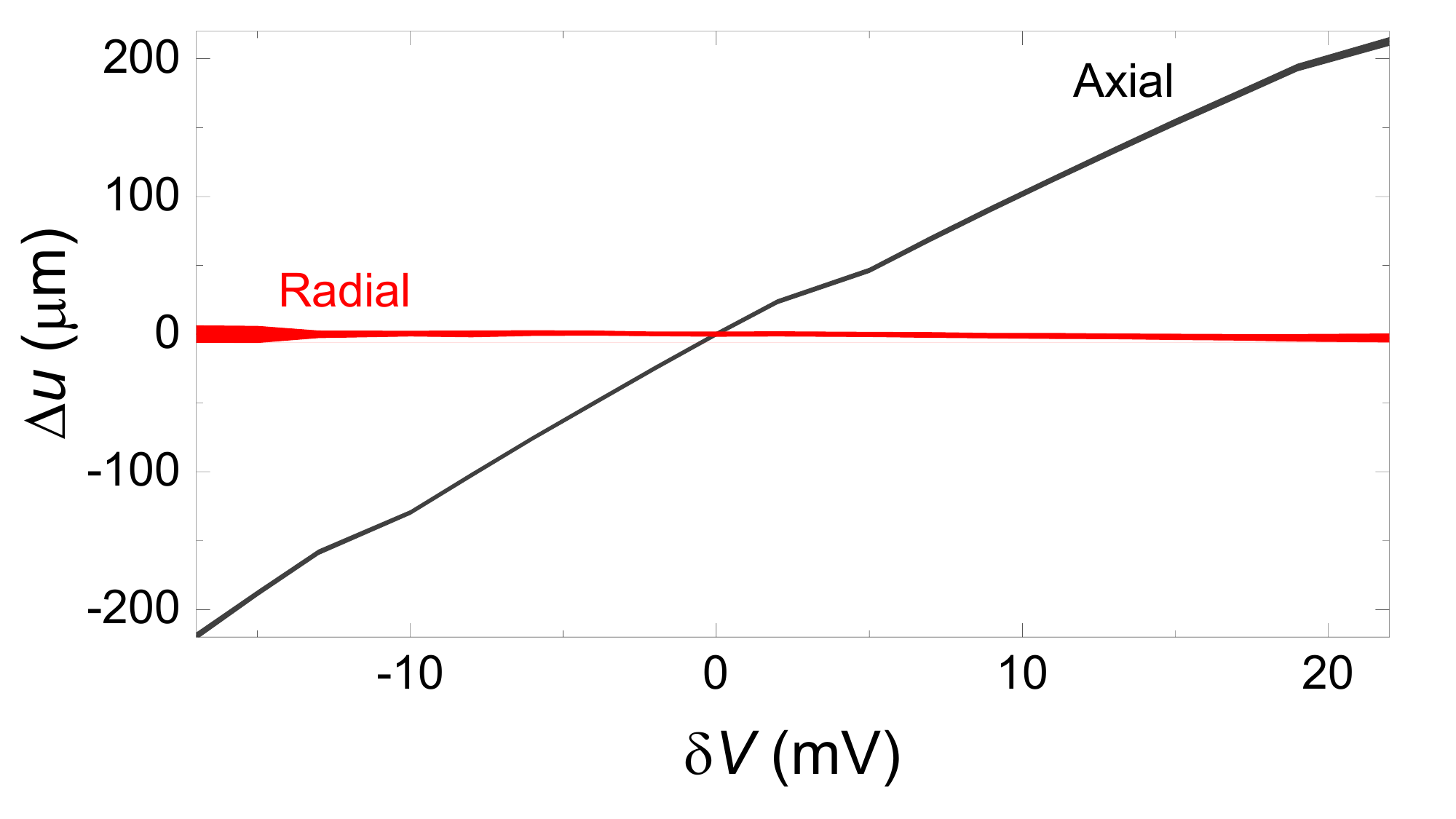}% Here is how to import EPS art
\vspace{-3mm}\caption{Position shift in the axial and radial directions as a function of the trap-depth difference ($\delta V$) relative to a symmetric trap potential configuration with a trap depth of 30~V, while the position of the radial laser beams remained fixed. The areas extend the values by $\pm 1$~$\sigma$. \label{fig:beam_alignment}}
\end{figure}

The alignment of the axial laser beams with the trap axis was verified by monitoring the fluorescence image of a single ion (Fig.~\ref{fig:excitation2}) while translating it to different axial positions within the trap. The ion displacement was achieved by introducing a controlled asymmetry in the trapping potential, with the corresponding trap-depth differences determined from simulations. The optimum alignment obtained using this procedure is shown in Fig.~\ref{fig:beam_alignment}, where one can see that $\sigma _z$ and $\sigma _r$ (shaded areas), obtained from the projections of the distribution of photons on the EMCCD camera on the axial and radial directions, remain the same within a few hundred $\mu$m. The images shown in Fig.~\ref{fig:excitation2} correspond to $\delta V=0$.

The performance of the MT is limited by deviations from the ideal homogeneous magnetic field, $\vec{B}$, and quadrupolar electric potential $\Phi$. 
While the correction electrode can be used to compensate higher-order electric-field contributions, residual distortions may remain owing to machining tolerances or imperfect alignment between the trap symmetry axis (dashed line in Fig.~\ref{fig:setup}(c)) and the magnetic-field axis (horizontal arrow). The misalignment corresponding to the measurements presented in this work is quantified in Appendix~\ref{app:misalignment}.

%\subsection{Axial laser-beam alignment}

\begin{table*}[t]
\caption{List of relevant Penning-trap parameters from the measurements with the new cryogen-free magnet. The table lists the average eigenfrequencies obtained from the measurements. For the trap depth of 30~V, measurements were carried out over seven days; however, only the daily average from one representative day (Fig.~\ref{fig:voltages}) is presented here. $\Delta \nu _c$ is the average of the difference between the cyclotron frequency obtained using Eq.~(\ref{eq:cyclotron_direct})~and~(\ref{eq:invariance_theorem}). The quantities $\sigma_z$ and $\sigma_r$ correspond to the projections of the fluorescence image recorded by the EMCCD camera from a single image (e.g. Fig.~\ref{fig:excitation2}(a)) onto the z- and r-directions, respectively. $d_\textrm{ion-ion}$ is obtained from Eq.~(\ref{eq:crystal_equilibrium_distance}). \label{tab:trap}}
\begin{ruledtabular}
\begin{tabular}{ccccccccc}
% &\multicolumn{2}{c}{$D_{4h}^1$}&\multicolumn{2}{c}{$D_{4h}^5$}\\
 Trap depth & $\langle\nu_z\rangle $ & $\langle\nu _-\rangle $ & $\langle\nu _+\rangle $
& $\langle 2(\nu_+\nu_-)/\nu_z^2 \rangle $ & $\langle \Delta \nu _c \rangle$ & $\sigma _z$ &$\sigma _r$& $d_\textrm{ion-ion}$\\ 
 (V) & (kHz) & (kHz) & (MHz) & & (Hz) &($\mu$m)& ($\mu$m)& ($\mu$m)\\
\hline
 10& 193.3546(7)& 6.97936(9)&  2.68311099(5)& 1.00178(1)& -12.40(7)& 1.67(1)& 2.40(1)& 16.5\\
 15& 236.9299(9)& 10.49154(19)& 2.67960220(10)& 1.00161(1)& -16.82(14)& 1.54(1)& 2.67(2)& 14.1\\
 20& 273.6779(15)& 14.01396(17)& 2.67608401(16)& 1.00143(1)& -19.81(9)& 1.45(1)& 1.84(1)& 13.3\\
 25& 306.0112(9)& 17.54389(12)& 2.67255807(17)& 1.00141(1)& -24.53(15)& 1.45(1)& 1.54(1)& 12.5\\
 30& 335.2835(10)& 21.08859(13)& 2.66901846(15)& 1.00139& -29.07(5)& 1.37(1)& 1.46(1)& 11.8\\
 %40& 386.770& 28.128& 2.661933& 1.00134& -37.3(4)& & & \\
\end{tabular}
\end{ruledtabular}
\end{table*}

\section{Single-Ion Photon-Based Detection}

The measurements presented in this work were carried out using two complementary photon-based detection techniques. The first is a pulsed optical method to determine the three eigenfrequencies that was introduced in Refs.~\cite{Berr2024,Berr2024_2} using the liquid-helium superconducting solenoid that was operated in our laboratory until December 2022. The second, referred to as Fluorescence-Detected Fourier-Transform Ion-Cyclotron Resonance (FD-FT-ICR), constitutes the optical analogue of the FT-ICR technique described in Refs.~\cite{Comi1974,Comi1974_2}, where the ion motion is detected through laser-induced fluorescence instead of the image current induced in the trap electrodes. In the former method, the cooling lasers are switched off during the motional excitation, whereas in FD-FT-ICR they remain on throughout the measurement.
\subsection{Measurements of all eigenfrequencies with the pulsed optical method: Fluorescence profile and photon counting}
The value of $B$ is obtained from $\nu_c$, which is determined with the invariance theorem (Eq.~\ref{eq:invariance_theorem}) from the three eigenfrequencies measured sequentially. Equation~(\ref{eq:cyclotron_direct}) can also be used although the value of $\nu_c$ will not be accurate due to the observed misalignment and ellipticity (see Appendix~\ref{app:misalignment}). Measurements performed for trap depths between 10~V and 40~V consistently showed that the cyclotron frequencies extracted from the invariance theorem were lower than those obtained from the sum of the radial eigenfrequencies. This is shown in Tab.~(\ref{tab:trap}) together with other important trap parameters.

\begin{figure}[t]
\includegraphics[scale=0.29]{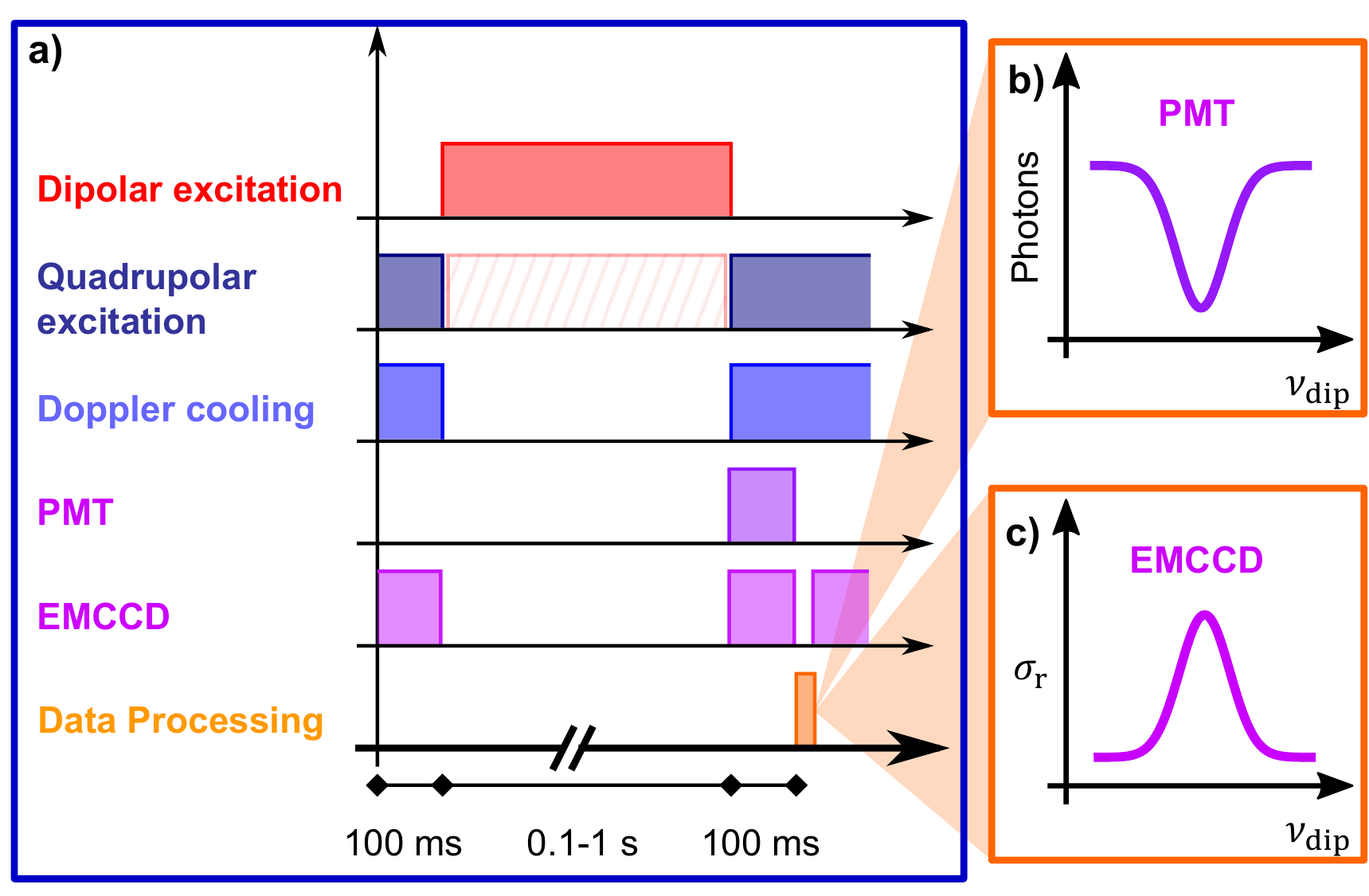}% Here is how to import EPS art
\vspace{-0mm}
%\caption{. \label{fig:fits}}
%\end{figure}

%\begin{figure}[h]
\includegraphics[scale=0.26]{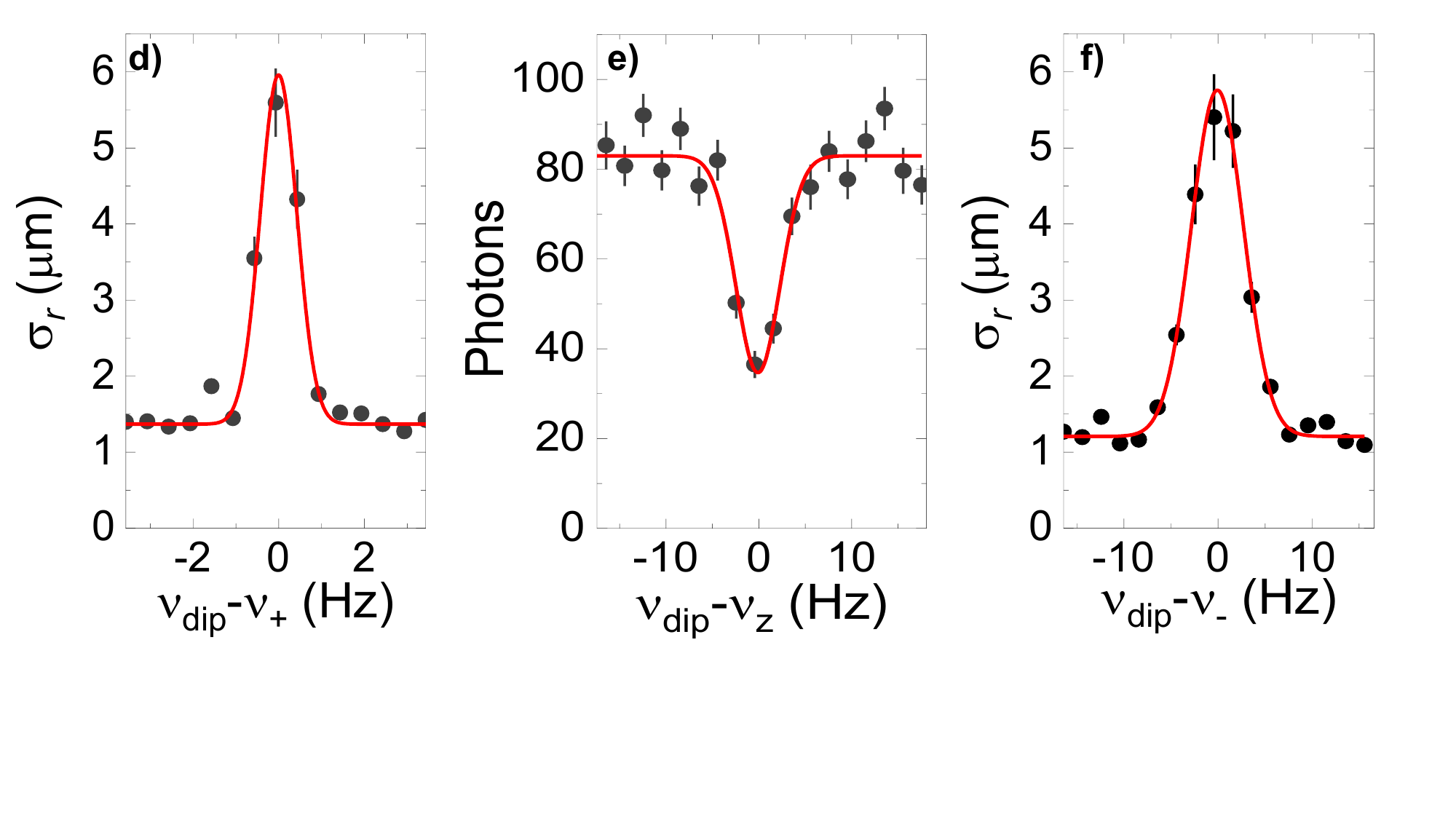}% Here is how to import EPS art
\vspace{-12mm}
\caption{Top: a) Measurement sequence used to determine any of the motional eigenfrequencies of Ca$^+$ from fluorescence emitted on the S$_{1/2}\rightarrow$P$_{1/2}$ transition (Fig.~\ref{fig:level_scheme}). (b) and c) Resonance profiles obtained with the PMT and the EMCCD camera, respectively. Bottom: Representative measurements of the d) modified-cyclotron, e) axial, and f) magnetron frequencies. The radial eigenfrequencies are extracted from images recorded with the EMCCD camera, whereas the axial frequency is measured using the PMT signal. More details can be found in Ref.~\cite{Berr2024_2}.  \label{fig:fits}}
\end{figure}

The sequence for one measurement scan of each of the eigenfrequencies is illustrated in Fig.~\ref{fig:fits}(a). It begins with Doppler cooling of the ion for approximately $100~\mathrm{ms}$. During this time, a quadrupolar radiofrequency field at approximately $\nu_c$ is applied for improved radial cooling (axialization \cite{Powe2002}). Subsequently, a dipolar radiofrequency excitation pulse is applied for a duration between $0.1$ and $1~\mathrm{s}$. After excitation, the ion is Doppler cooled again, and the fluorescence signal is detected using the PMT and the EMCCD camera. This sequence is repeated for different values of $\nu_{\mathrm{dip}}$ across the eigenfrequency. The PMT measurement records the fluorescence photons as a function of $\nu_{\mathrm{dip}}$, producing a dip in the photon counts at resonance (Fig.~\ref{fig:fits}(b)). This is due to an increase in the oscillation amplitude, resulting in fewer photons being detected within the first milliseconds. The EMCCD measurement determines the spatial width $\sigma_r$ of the ion image, which increases near resonance as a result of the excitation of any of the radial motions (Fig.~\ref{fig:fits}(c)). Further details of the method can be found in Ref.~\cite{Berr2024,Berr2024_2}. %\textcolor{red}{\bf Further details are given in Appendix~\ref{app:optical_method}, extracted from Ref.~\cite{Berr2024_2}.}

The optimal excitation time was different for each eigenfrequency ($\nu_+,\nu_z,\nu_-$), since the propagation of their uncertainties in $\nu_c$ is given by
\begin{equation}
\sigma _{\nu_c} =\sqrt{
\left( \frac{\nu_+}{\nu_c} \sigma _{\nu_+} \right)^2 +
\left( \frac{\nu_z}{\nu_c} \sigma _{\nu_z} \right)^2 +
\left( \frac{\nu_-}{\nu_c} \sigma _{\nu_-} \right)^2}.
\end{equation}

Measurements performed with different excitation times require different acquisition times, amplitudes, and numbers of repetitions. In order to compare different excitation times directly, the statistical improvement associated with the number of repetitions was removed by defining a normalized uncertainty \cite{Berr2024_2}
\begin{equation}
\Delta \nu_u = \sigma _{\nu_u}\sqrt{\frac{t^u_{\mathrm{dip}}}{t_0}\frac{N}{N_0}},
\end{equation}
where $t^u_{\mathrm{dip}}$ is the excitation time, $N$ is the number of repetitions, and $t_0$ and $N_0$ are reference values. The factor $\sqrt{N/N_0}$ appears because the experimentally measured uncertainty already contains the $1/\sqrt{N}$ statistical improvement; multiplying by $\sqrt{N}$ removes this dependence and allows comparisons between scans performed with different acquisition times and repetition numbers. The quantity $\Delta \nu_u$ therefore represents the frequency uncertainty achievable for a fixed total experimental duration. Using this normalization, the optimal excitation times were fixed to $1000$ ms for $\nu_+$, $100$ ms for $\nu_-$, and $100$ ms for $\nu_z$ for the measurements presented here. 

%The cyclotron frequency $\nu _c$ has been determined with the invariance theorem (Eq.~\ref{eq:invariance_theorem}), 

\begin{figure}[t]
\includegraphics[scale=0.46]{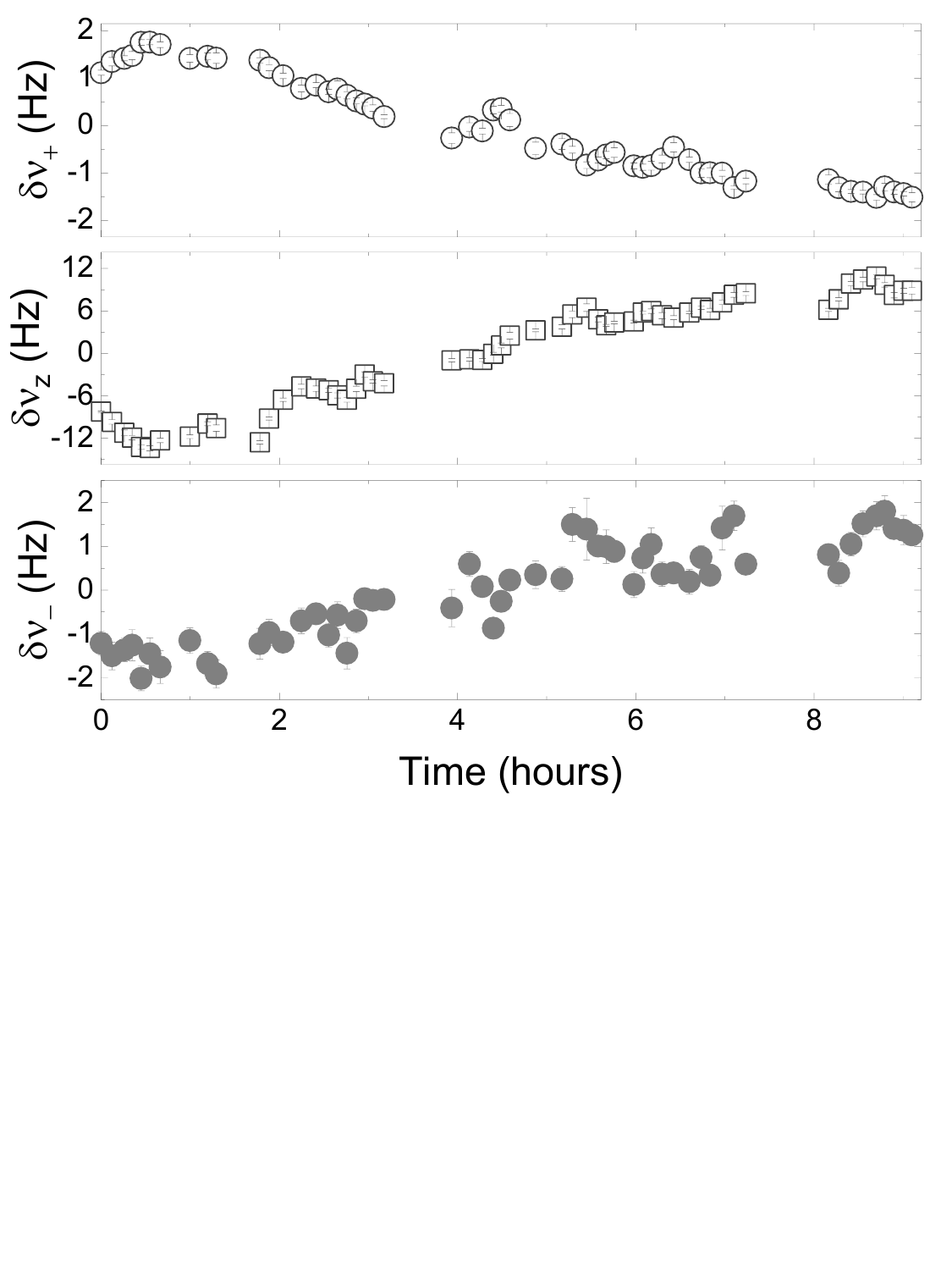}% Here is how to import EPS art
\vspace{-45mm}\caption{Deviation $\delta \nu_u$ of the eigenfrequencies $\nu_u$ from their respective mean values $\langle \nu_u \rangle$ on a one-day measurement. Top: Reduced-cyclotron frequency with $\langle \nu_+ \rangle = 2669.0186$~kHz. Middle: Axial frequency with $\langle \nu_z \rangle = 335.2838$~kHz. Bottom: Magnetron frequency with $\langle \nu_- \rangle = 21.0886$~kHz. 
\label{fig:voltages}}
\end{figure}

\begin{figure}[h]
\includegraphics[scale=0.33]{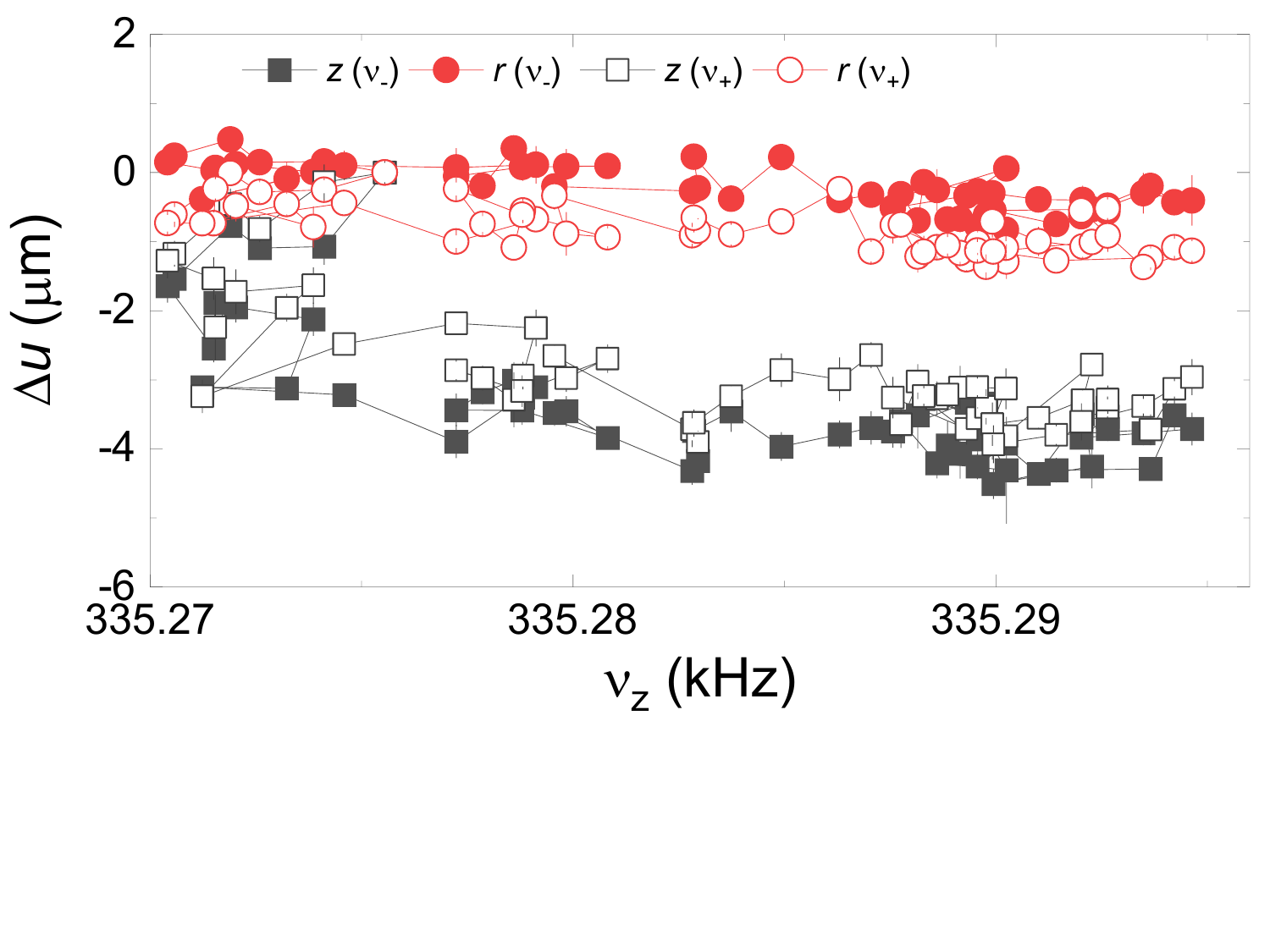}% Here is how to import EPS art
\vspace{-18mm}\caption{Deviation of the ion position from its initial value as a function of the axial frequency for the measurements presented in Fig.~\ref{fig:voltages}. The coordinates in the radial plane (circles) and in the axial direction (squares) are obtained from images of the radial eigenmotions used for the frequency measurements. \label{fig:position_2}}
\end{figure}

The lower panels of Fig.~\ref{fig:fits} display resonances corresponding to the three eigenmotions, each of them built from several scans. Figure~\ref{fig:voltages} presents the results obtained over the course of one day. The axial frequency depends exclusively on the DC voltage applied to the trap electrodes. The drift observed throughout the day also affects the radial eigenfrequency values. Figure~\ref{fig:position_2} shows the ion's position in the radial and axial direction as a function of the axial frequency for the measurements presented in Fig.~\ref{fig:voltages}. The ion's radial position remains within less than a micrometer, while the axial position changes slightly due to minimal changes in the potential well (see also Fig.~\ref{fig:beam_alignment}).

The final value of each eigenfrequency is the weighted mean of the individual eigenfrequency measurements from the fit in each scan,
\begin{equation}
\bar{\nu} =\frac{\sum_{i=1}^{N} w_i  \nu_i}{\sum_{i=1}^{N} w_i},
\end{equation}
where $w_i = 1/(\Delta \nu_i)^2$. Two uncertainties can be associated with the sample \cite{Birg1934}. The internal uncertainty
\begin{equation}
\Delta \bar{\nu}_{\mathrm{int}} =\sqrt{\frac{1}{\sum w_i}},
\end{equation}
calculated from Gaussian error propagation, and the external uncertainty
\begin{equation}
\Delta \bar{\nu}_{\mathrm{ext}} =\sqrt{\frac{\sum w_i (\nu_i - \bar{\nu})^2}{(N-1)\sum w_i}},
\end{equation}
which accounts for the statistical dispersion of the sample $\nu_i$.
%If the uncertainties are well estimated for each $\nu_i$, $\Delta \bar{\nu}_{\mathrm{int}}$ is a good estimate of the eigenfrequency uncertainty. 
The eigenfrequency uncertainty is chosen as
\begin{equation}
\Delta \bar{\nu} = \max(\Delta \bar{\nu}_{\mathrm{int}}, \Delta \bar{\nu}_{\mathrm{ext}}).
\end{equation}

This is the only uncertainty considered for the measurements presented here. %which gathers systematic effects.

\subsection{Measurements with the FD-FT-ICR technique}

The PD-FT-ICR technique, presented here for the first time, applies the FFT to the time stamp of photons recorded with the PMT, while probing the modified-cyclotron motion with near-resonant external drives. Owing to the weak interaction between the radial laser beams and the ion, together with the laser beam profile, the reduced-cyclotron frequency can be detected in addition to the frequency of the drive \cite{DeMiguel2026}. For each data file, the arrival times were first binned into a histogram over a total acquisition window of $T = 25~\mathrm{s}$, using a bin size of $\Delta t = 160~\mathrm{ns}$.
This corresponds to a sampling frequency $f_s = 6.25$~MHz and to a Fourier frequency resolution $\Delta f = 0.04~\mathrm{Hz}$. The 25~s histogram was divided into five consecutive 5~s intervals. Each interval was then embedded into a 25~s array by padding the remaining bins with zeros. The FFT was subsequently calculated from this zero-padded histogram. The resulting spectrum for the $k$-th interval is denoted by
\begin{equation}
I_k(f)=\left|\mathrm{FFT}\left[ h_k^{(\mathrm{zp})}\right]\right|,
\end{equation}
where $h_k^{(\mathrm{zp})}$ is the histogram corresponding to the $k$-th 5~s interval after zero-padding. For each input file, the five FFT spectra were then combined to obtain the mean spectrum.
%\begin{equation}
%I_{\mathrm{mean}}(f)=\frac{1}{5}\sum_{k=1}^{5} I_k(f).
%\end{equation}
\begin{figure}[t]
\includegraphics[scale=0.33]{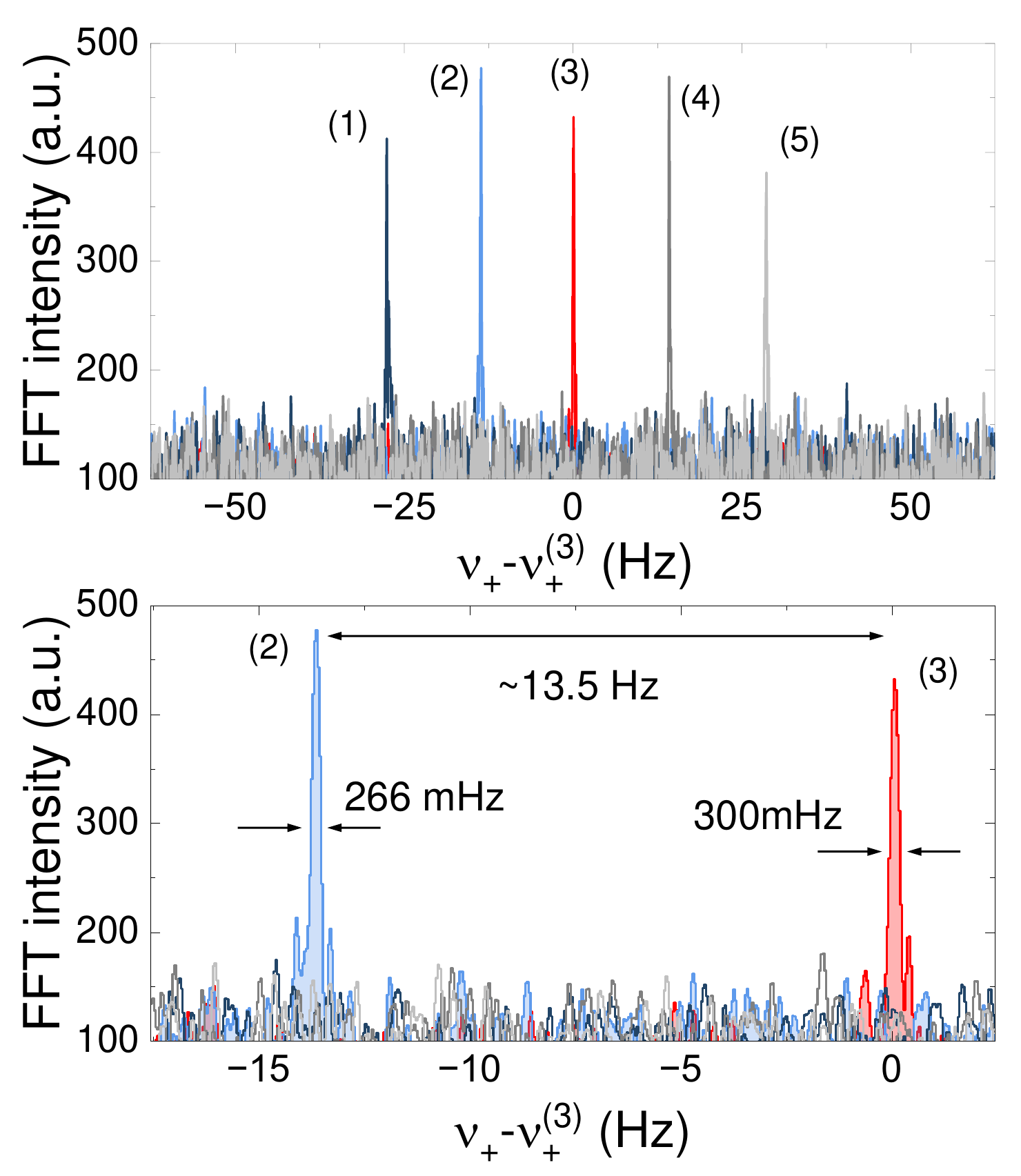}% Here is how to import EPS art
\vspace{-3mm}\caption{Upper panel: FFT spectrum obtained for different trap depths separated by only 0.02~V, corresponding to the minimum resolution of the power supply ((1)–(5)). Lower panel: Zoom of the frequency region around peaks (2) and (3), illustrating the high resolution of the method. The full width at half maximum is indicated. \label{fig:FFT}}
\end{figure}

The upper panel of Fig.~\ref{fig:FFT} shows the FFT spectrum around $\nu_+$ when applying a radiofrequency-comb excitation for five different trap depths. The comb is about 300~Hz away and not visible in the figure. Over a trapping potential of $V_0 = 30~\mathrm{V}$, the minimum power supply resolution of $0.02$~V corresponds 
%Consecutive trap depths differ by 0.02~V, corresponding to the minimum voltage increment achievable with the laboratory power supply. This, over a trapping potential of $V_0 = 30~\mathrm{V}$ corresponds 
to a relative variation of $6.67\times 10^{-4}$. Using Eq.~(\ref{eq:redcyc_mag}), this leads to 
\begin{equation}
\frac{\Delta \nu_+^*}{\nu_+} \simeq -\frac{\nu_-}{\nu_+-\nu_-} \frac{\Delta V_0}{V_0},\label{eq:resolution1}
\end{equation}

where $\Delta \nu_+^*$ denotes the frequency shift associated with the minimum voltage step of the power supply. While the frequencies of the comb components remain fixed, the reduced-cyclotron frequency shifts as a result of the change in trap depth. This effect is highlighted in the lower panel of Fig.~\ref{fig:FFT}, which presents a zoomed view for two consecutive trap depths of 29.98 and 30.00~V. The observed frequency difference of $|\Delta \nu_+^*| \approx 13.5~\mathrm{Hz}$ is in agreement with the value expected from the trap-depth variation. 

The mass resolving power is determined independently from the linewidth of the FFT peak. From the lower panel of Fig.~\ref{fig:FFT} the measured full width at half maximum (FWHM) at $V_0=30$~V, obtained from a 25-s acquisition, yields
\begin{equation}
R_{(m/q)}\approx \frac{\nu_+}{\Delta \nu_+(\hbox{FWHM})} = 8.9\times10^{6}.
\end{equation}

In the context of this paper this method provides measurements within constant time intervals of five seconds. Assuming that the linewidth is limited by the observation time and therefore scales inversely with the acquisition time, the corresponding mass resolving power for a 5-s measurement is estimated to be $R_{(m/q)}\simeq 1.8\times10^{6}$.

\section{Magnetic field stability}

\begin{figure*}[t]
\includegraphics[scale=0.45]{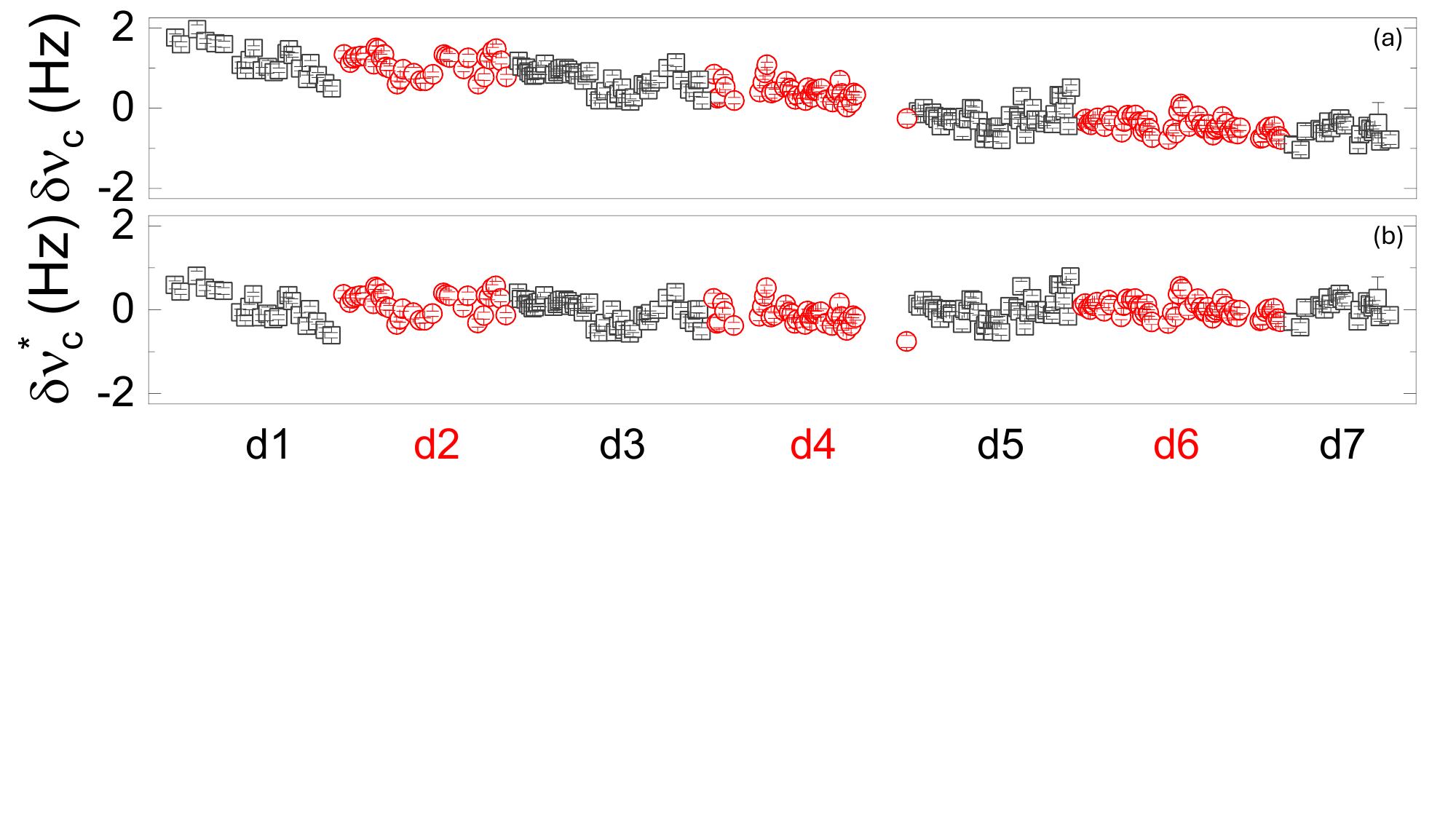}% Here is how to import EPS art
\vspace{-38mm}\caption{(a) Deviation of the cyclotron frequency from the reference value $\nu_c = 2.690107.25$~Hz over seven days of measurements (d1--d7), with the cyclotron frequency evaluated using the invariance theorem, Eq.~(\ref{eq:invariance_theorem}). (b) Deviation of the cyclotron frequency after subtraction of the long-term magnetic-field drift from the data shown in panel (a). The periods during which no data were acquired are omitted for better visualization. See text for details. \label{fig:long_term_drift}}
\end{figure*}
The magnetic-field stability was evaluated using the two methods introduced in the previous section. Figures~\ref{fig:long_term_drift}(a) shows the cyclotron-frequency variations obtained from the three eigenfrequencies using the pulsed optical method over four consecutive days, followed by a three-day interruption, and a further three consecutive days of measurements. Data collection was restricted to daytime hours. For clarity, the figure presents the measurement days as contiguous, omitting the periods during which no data were acquired. Equation~(\ref{eq:invariance_theorem}) was applied to determine the frequencies shown in panel (a). For instance, the measurements recorded on day six (d6) were calculated from the eigenfrequency values presented in Fig.~\ref{fig:voltages}. Because the three eigenfrequencies were measured sequentially, the acquisition of each data point in Fig.~\ref{fig:long_term_drift} required several minutes. The axial and magnetron frequencies each took approximately 1.5~min to measure, while the modified-cyclotron frequency measurement required about 2~min. To assign a common timestamp to each measurement block (or data point) in Fig.~\ref{fig:long_term_drift}, a reference time, $t_{\mathrm{lin}}$, was defined as the midpoint between the start of the axial-frequency measurement, and the end of the reduced-cyclotron-frequency measurement.
%\begin{equation}
%t_{\mathrm{lin}} = \frac{t_{0} + t_{1}}{2}.
%\end{equation}
The result is a triplet $\{\nu_{z}^{\mathrm{lin}},\, \nu_{-}^{\mathrm{lin}},\, \nu_{+}^{\mathrm{lin}}\}$ at $t_{\mathrm{lin}}$. The associated uncertainty to $\nu _c$ is 
\begin{equation}
\sigma_{\nu_c^{\mathrm{lin}}}=
\frac{\sqrt{\left(\nu_z^{\mathrm{lin}}\sigma_{\nu_z^{\mathrm{lin}}}\right)^2
+
\left(\nu_-^{\mathrm{lin}}\sigma_{\nu_-^{\mathrm{lin}}}\right)^2
+
\left(\nu_+^{\mathrm{lin}}\sigma_{\nu_+^{\mathrm{lin}}}\right)^2}
}{\nu_c^{\mathrm{lin}}}.
\end{equation}
A weighted least-squares regression of $\nu_c$ as a function of time was performed using the individual measurement uncertainties as statistical weights, resulting in
\begin{equation}
\frac{d\nu_c}{dt}=-2.29(24)\times10^{-6}\,\mathrm{Hz/s},
\end{equation}
where the quoted uncertainty corresponds to the conservative block-bootstrap estimate. This result is further supported by the 95\% confidence interval, $\left[-3.22,\,-1.99\right]\times10^{-6}\,\mathrm{Hz/s}$, which excludes a zero slope by a wide margin. Consistent with this finding, the corresponding hypothesis test yields a $p$-value below $10^{-15}$, providing strong evidence for a non-zero linear drift. From the coefficient of determination, $R^2=0.84$, the linear drift accounts for 84\% of the total variance observed in the cyclotron-frequency data. Expressed in relative terms, the long-term drift corresponds to
\begin{equation}
\frac{1}{B} \frac{dB}{dt} = -3.07(32)\times10^{-9}\,\mathrm{h}^{-1},
\end{equation}
which is of the same order of magnitude as values reported for other liquid-helium-cooled superconducting magnets used in precision Penning-trap experiments on exotic nuclei, where pressure and temperature stabilization has been implemented~\cite{Mari2008}. %A detailed comparison with literature values and a discussion of possible environmental influences are provided in Sec.~4.6.

Although the linear trend accurately captures the long-term behavior of the magnetic field, the reduced chi-squared statistics, $\chi_\nu^2 = 13.7$ $(\mathrm{DoF}=257)$, is significantly larger than unity. Since $\chi_\nu^2 \approx 1$ is expected when the residual scatter is fully explained by the stated uncertainties, this elevated value indicates the presence of additional fluctuations beyond the long-term drift. To investigate the origin of these short-term fluctuations, several studies were performed, beginning with the stability of $V_0$. Considering a representative measurement day, such as d6 (Fig.~\ref{fig:voltages}), the axial frequency varies by 24.2~Hz over the course of the day, whereas the magnetron and reduced-cyclotron frequencies fluctuate by only about 3.3~Hz. The axial frequency scales with the square root of $V_0$ (Eq.~\ref{eq:axial_frequency}) and, for a trap with only small misalignments, it can be considered essentially independent of the magnetic field. Therefore, the observed drift of the axial frequency must arise from fluctuations of $V_0$, i.e., from instabilities of the power supply. Although the absolute value of $V_0$ may vary between successive eigenfrequency triplets without affecting the reconstructed cyclotron frequency, it must remain constant during the acquisition of an individual triplet. In practice, this condition is not fully satisfied. For a trap depth of 30~V, voltage drifts of up to $0.59~\mathrm{mV/h}$, corresponding to $19.6~\mathrm{ppm/h}$, were observed, which are large compared to the magnetic field stability of 0.1~ppm. The possible influence of these voltage fluctuations on the reconstructed cyclotron frequency was therefore investigated in detail. Although a measurable correction is obtained, it does not modify the inferred magnetic-field stability obtained after linearization. The corresponding analysis is presented in Appendix~\ref{app:newton_raphson}.

For further investigations of short-term fluctuations of the magnetic field, the long-term linear drift was subtracted from the data presented in Fig.~\ref{fig:long_term_drift}(a), leaving only the residuals (Fig.~\ref{fig:long_term_drift}(b)). A quantitative comparison of the residual scatter, $\sigma_s = 0.278$~Hz, with the root-mean-square measurement uncertainty, $\sigma_{\mathrm{rms}} = 0.091$~Hz, yields a ratio of $\sigma_s/\sigma_{\mathrm{rms}} = 3.07$, demonstrating the presence of fluctuations beyond the quoted measurement uncertainties. The conclusions are presented in Appendix~\ref{sec:temp_magnetic_field}, although this is limited because of the measurement time. 

Measurements using the FD-FT-ICR technique were carried out to overcome this limitation.
\begin{figure}[t]
\includegraphics[scale=0.45]{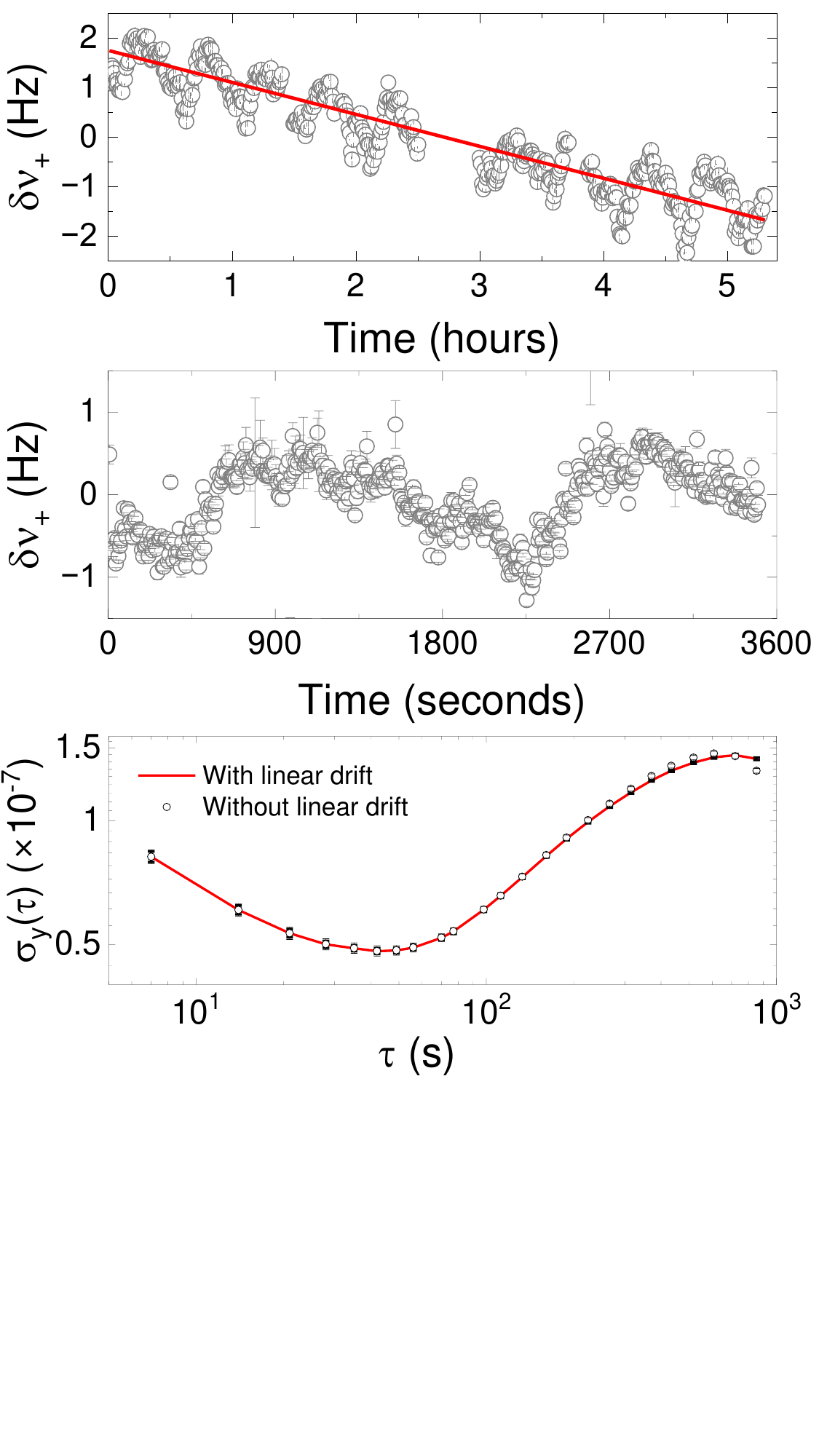}% Here is how to import EPS art
\vspace{-43mm}\caption{Upper panel: Measurement of $\nu_+$ over a full day using the fluorescence-photon modulation technique. Each data point corresponds to 25~s acquisition. Middle panel: First hour of the dataset shown in the upper panel. Each data point corresponds to 5~s acquisition. Lower panel: Overlapping Allan deviation of the data shown in the middle panel, calculated with and without removal of the long-term magnetic-field drift. See text for details. \label{fig:FFT_plot}}
\end{figure}
The evaluation of the short-term magnetic-field stability was done through the overlapping Allan deviation of the measured reduced-cyclotron frequency of a single $^{44}$Ca$^+$ ion shown in the middle panel of Fig.~\ref{fig:FFT_plot}, calculated for a set of averaging times ($\tau=m\Delta t$). The upper panel of Fig.~\ref{fig:FFT_plot} shows the full time evolution of the measured $\nu_+$ fluctuations over approximately five hours. A clear slow structure is observed, indicating that the magnetic-field variations are not dominated only by uncorrelated white noise. The middle panel corresponds to the first hour of the same dataset and therefore represents a short-time zoom of the initial part of this evolution. The reference frequency ($\nu_+^0=2424987.15$~Hz) is calculated as the weighted mean of the dataset ($\nu_+^i$) and the frequency fluctuations are expressed as fractional deviations,
\begin{equation}
y_i=\frac{\nu_+^i-\nu_+^0}{\nu_+^0}.
\end{equation}
The overlapping Allan deviation was calculated as
\begin{equation}
\sigma_y(\tau)=\sqrt{\frac{1}{2(N-2m+1)}\sum_{k=1}^{N-2m+1}\left(\bar y_{k+1}-\bar y_k\right)^2},
\end{equation}
where 
\begin{equation}
\bar y_k
=
\frac{1}{m}
\sum_{j=k}^{k+m-1} y_j
\end{equation}
is the average fractional frequency over a window containing $m$ consecutive measurements. The resulting $\sigma _y (\tau)$ calculated with and without removing the long-term drift visible in the full data set is shown in the lower panel of Fig.~\ref{fig:FFT_plot}. %The uncertainties were estimated using a Monte Carlo procedure in which the individual frequency measurements were randomly sampled according to their fitted uncertainties. The Allan deviation was then recalculated for each realization. 
Assuming that $\nu_+ \propto B$, the obtained Allan deviation directly reflects the relative magnetic-field stability,
\begin{equation}
\frac{\delta B}{B}=\frac{\delta \nu_+}{\nu_+}.
\end{equation}
The Allan deviation decreases with increasing averaging time and reaches a minimum value of approximately $5\times 10^{-8}$ for averaging times between 30 and 60~s. This corresponds to a relative magnetic-field stability of $5\times 10^{-8}$. For the measured reduced-cyclotron frequency $\nu_+^0=2424987.15$~Hz, the minimum Allan deviation corresponds to an absolute frequency stability of approximately 0.12~Hz. At longer averaging times, the Allan deviation increases and reaches $\sim 1.4 \times 10^{-7}$ at $\tau \approx 10^3$~s, indicating the presence of correlated magnetic-field fluctuations on time scales of several hundred seconds. The nearly identical Allan deviations obtained before and after subtraction of the linear drift suggest that these long-term fluctuations are not dominated by a simple linear drift, but rather by low-frequency correlated variations which are clearly periodic and are probably assigned to temperature and/or pressure fluctuations.
\begin{figure}[t]
\includegraphics[scale=0.26]{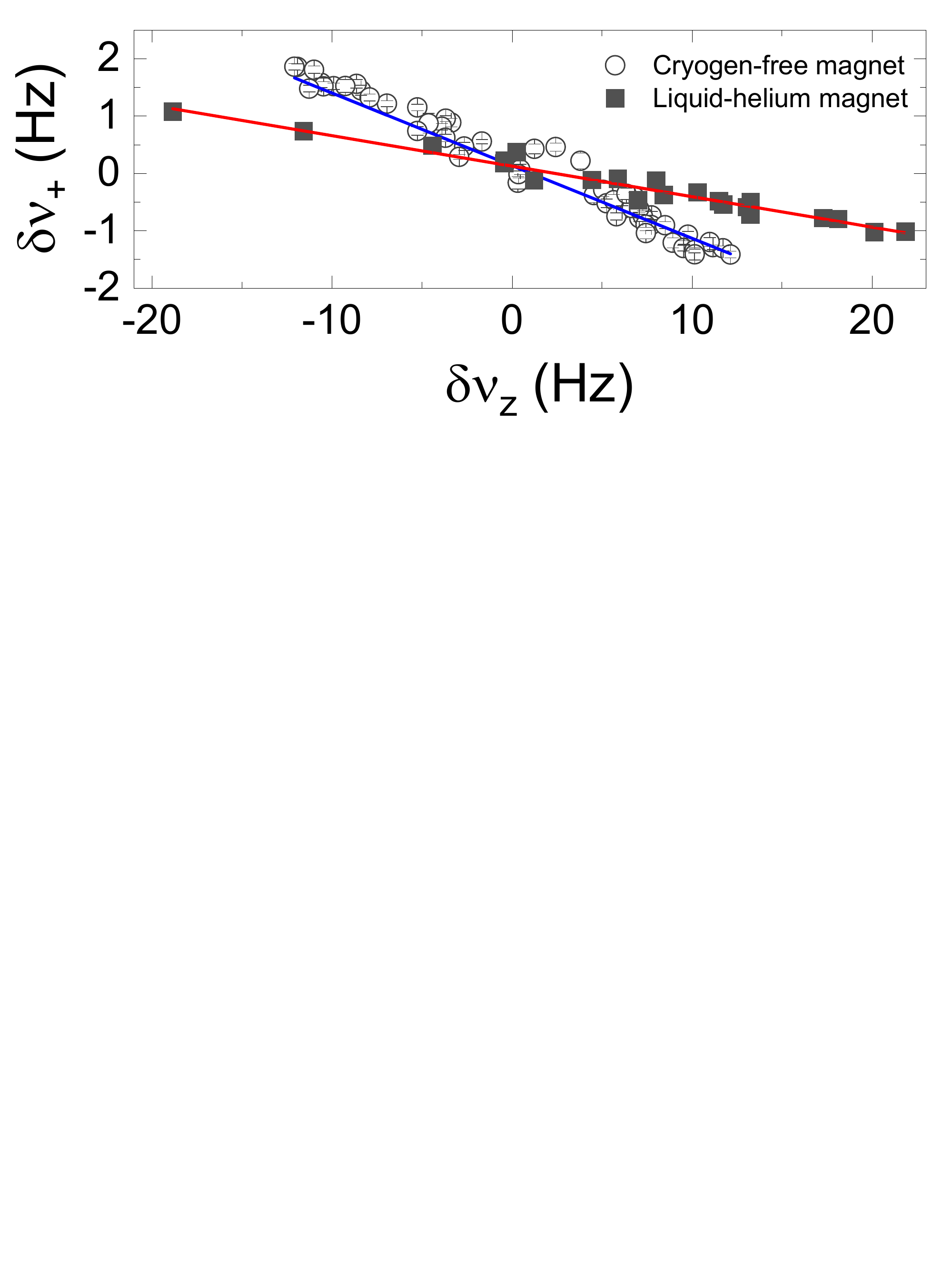}% Here is how to import EPS art
\vspace{-80mm}\caption{Variation of $\nu_+$ as a function of $\nu_z$ for measurements performed with two different magnets using the pulsed optical method. The solid lines show fits based on Eq.~(\ref{eq:redcyc_mag}). The differing slopes reflect the different $\nu_z$ values corresponding to different trap depths. See the text for further discussion.\label{fig:magnet}}
\end{figure}

Finally, a direct comparison between the present cryogen-free magnet and a liquid-helium-based system could be established using previous measurements performed in the laboratory with the pulsed optical technique. Representative results are shown in Fig.~\ref{fig:magnet}. A total of 22 data points were collected over four days using the liquid-helium magnet before decommissioning, whereas 55 data points were acquired within a single day using the cryogen-free magnet (d6 in Fig.~\ref{fig:long_term_drift}). Deviations of the data points from a linear fit are expected to arise primarily from magnetic-field fluctuations and other sources of experimental noise. The corresponding standard deviations of the residuals amount to $\sigma_d = 0.21$ and $\sigma_d = 0.11$ for the cryogen-free and liquid-helium magnets, respectively. The coefficients of determination obtained from the linear fits are $R^2=0.9753$ and  $R^2=0.9596$. Despite the larger deviations observed for the cryogen-free magnet, its higher $R^2$ value indicates a stronger linear correlation between the eigenfrequency variations, reflecting the larger overall range of the measured frequency shifts. As expected, magnetic-field-induced fluctuations are larger in the cryogen-free system than in the liquid-helium-based magnet. This difference is expected to be reduced through further mitigation of vibration-induced disturbances, for example by employing a more massive support structure. In neither case was the magnet actively stabilized. 
\section{Conclusions and Outlook}
In this paper, a detailed study of the magnetic-field stability of a Penning trap operated with a cryogen-free superconducting magnet using a single laser-cooled ion as sensor has been carried out. The magnet is cooled by a GM cryocooler, which introduces mechanical vibrations whose amplitudes and frequency spectrum have been characterized using standard diagnostic techniques. The vacuum chamber housing the trap is mechanically decoupled from the magnet in order to mitigate their impact, although residual vibrations may still affect the experimental performance.

The first photon-based detection technique developed within our group (pulsed optical method) \cite{Berr2024}, requiring between 5 and 10 minutes for a single cyclotron-frequency determination, has been employed to investigate the long-term magnetic-field stability. The relative drift measured in this work, $-3.07(32)$~ppb/h, is very similar to values reported by other high-precision Penning-trap experiments employing liquid-helium-based superconducting magnets. In particular, the ISOLTRAP experiment at CERN, operating at a magnetic field strength of approximately 5.9~T, reported a relative drift of $-2.9(6)$~ppb/h~\cite{Mari2008}, while SHIPTRAP measured a value of about $-1.3$~ppb/h~\cite{Giac2025}. Lower drift rates have been achieved at experiments such as PENTATRAP at the Max Planck Institute for Nuclear Physics in Heidelberg, where a value of $-0.23$~ppb/h was reported at a magnetic field strength of approximately 7~T~\cite{Krom2022}. However, this level of performance relies on extensive stabilization measures, including active temperature regulation of the laboratory environment to better than 0.1~K/day, as well as stabilization of the liquid-helium level and helium gas pressure within the magnet bore.

Envisaged spectroscopy experiments on thorium and thorium oxide ions require a level of magnetic-field stability compatible with single-ion identification at high sensitivity and precision. The FD-FT-ICR technique has demonstrated a relative magnetic-field stability $\delta B/B$ better than $5\times10^{-8}$ for averaging times below one minute. Using the same method, a mass-to-charge resolving power of $R \simeq 8.9\times10^{6}$ has been achieved for $^{40}$Ca$^+$ in 7~T. Even shorter measurement times are possible, with successful modified-cyclotron-frequency determinations obtained within a time window of 5~s, as shown in Fig.~\ref{fig:FFT_plot}(b), and with realistic prospects of reaching the 1~s regime. To the best of our knowledge, no previous Penning-trap experiment has reported the use of direct modified-cyclotron-frequency measurements on such short timescales for the purpose of monitoring magnetic-field stability. In most high-precision Penning-trap facilities, magnetic-field variations are inferred from periodic reference-ion measurements separated by substantially longer time intervals and are primarily used to correct magnetic-field drifts rather than to directly resolve short-term fluctuations.

The extension of the technique towards higher-precision measurements will require the implementation of active stabilization of environmental and magnet-related parameters, together with further mitigation of vibration-induced disturbances. On the one hand, vibration-induced magnetic-field fluctuations have previously been identified as a significant limitation in superconducting magnet systems~\cite{Brit2016}. This suggests that part of the residual short-term fluctuations observed in the present cryogen-free setup  may originate from mechanical vibrations associated with the cryocooler and supporting structure. On the other hand, periodic fluctuations similar to those shown in Fig.~\ref{fig:FFT_plot} have been observed in another experiment in the laboratory with the magnet compressor witched off, and are most likely linked to temperature fluctuations.

\begin{acknowledgments}
We warmly thank Roberto Palma and Rafael Gallego for their support on the mechanical vibration measurements. We acknowledge support from Grant No. PID2022-141496NB-I00 funded by MCIU/AEI /10.13039/501100011033 and by ERDF, EU,  by the infrastructure project No. IE19-204 UGR (funded by Junta de Andaluc\'ia/FEDER),
%Grant No. PID2019-104093GB-I00 funded by MCIU/AEI /10.13039/501100011033, from FEDER/Junta de Andalucía - Consejería de Universidad, Investigación e Innovación through Project No. P18-FR-3432, 
from Programa ``Yo Investigo" Junta de Andaluc\'ia-Next Generation EU, and from the University of Granada ``Laboratorios Singulares 2020".  %The construction of the facility was supported by the European Research Council (Contract No. 278648-TRAPSENSOR), Projects No. FPA2015-67694-P (funded by MCIU/AEI/10.13039/501100011033 and by ERDF A way of making Europe) and No. FPA2012-32076 (MCIU/FEDER), infrastructure Projects No. UNGR10-1E-501, and No. UNGR13-1E-1830 (MCIU/FEDER/UGR), and No. EQC2018-005130-P (funded by MCIU/AEI/10.13039/501100011033 and by ERDF A way of making Europe), and infrastructure Projects No. INF-2011-57131 and No. IE2017-5513 (funded by Junta de Andalucía/FEDER).
\end{acknowledgments}
\appendix
\section{Effect of misalignment}
\label{app:misalignment}
\begin{figure}[t]
\includegraphics[scale=0.25]{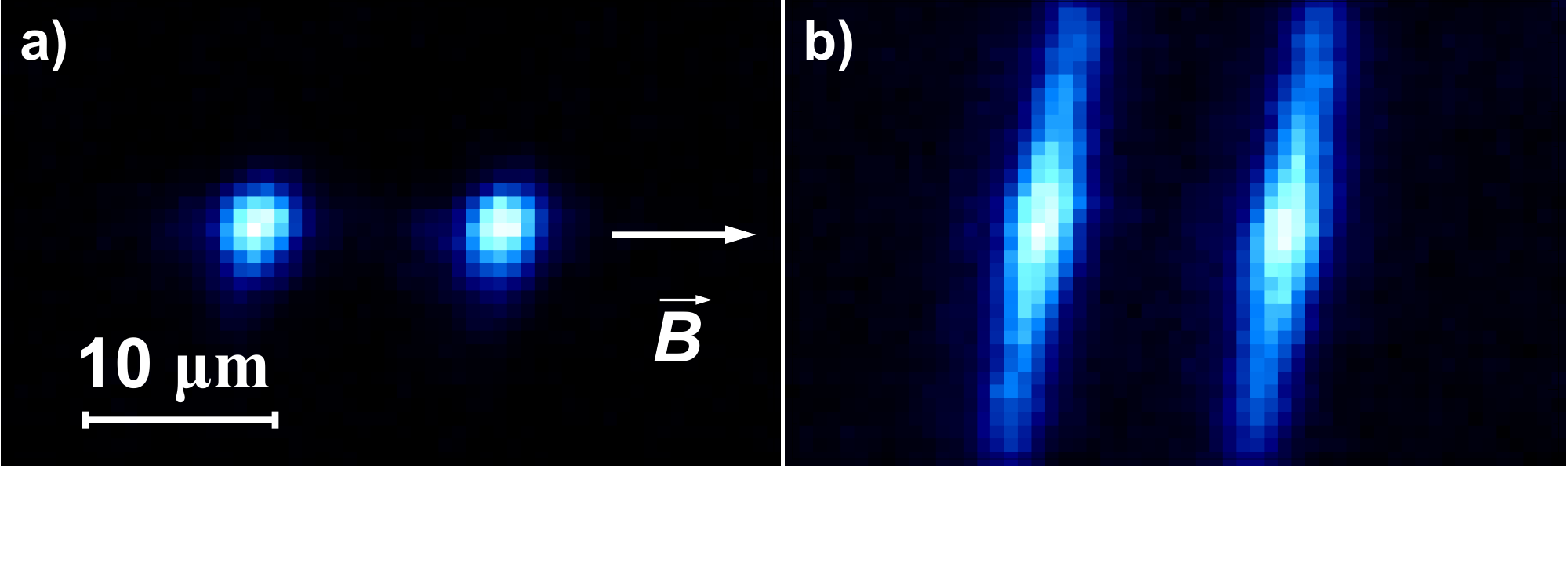}% Here is how to import EPS art
\vspace{-7mm}\caption{a): Image of a balanced Coulomb crystal. 
b): Image of the same crystal after applying a near-resonant excitation to the modified-cyclotron motion. \label{fig:excitation}}
\end{figure}

The effect of misalignment can be described in terms of $\theta_{\mathrm{trap}}$ (Fig.~\ref{fig:setup}(c)) and the ellipticity parameter $\epsilon_{\mathrm{trap}}$, which are related by~\cite{Gabr2009}
\begin{equation}
\frac{9}{4}\theta_{\mathrm{trap}}^2-\frac{1}{2}\epsilon_{\mathrm{trap}}^2
\approx
\frac{2\nu_+\nu_-}{\nu_z^2}-1.
\label{eq:misalignment_correlation}
\end{equation}
Analysis of the EMCCD images shown in Fig.~\ref{fig:excitation} revealed a coupling between the modified-cyclotron and axial motion. In the absence of excitation, the Coulomb crystal is aligned along the trap axis.  However, when the modified-cyclotron radius is increased through resonant excitation, the fluorescence distribution is no longer confined to the direction perpendicular to the trap axis but instead appears tilted by an angle $\theta_{\mathrm{obs}}$ with respect to it. This behavior is consistent with a misalignment between the electric- and magnetic-field axes. Nevertheless, a direct correspondence between $\theta_{\mathrm{obs}}$ and the actual trap misalignment angle, $\theta_{\mathrm{trap}}$, cannot be assumed.

The angle in Fig.~\ref{fig:excitation} was determined from 10 independent EMCCD images containing Coulomb crystals with more than two ions, yielding an average value of $\bar{\theta}_{\mathrm{obs}}=4.91(51)^\circ$ \cite{Schu2026}. Substituting this value into Eq.~(\ref{eq:misalignment_correlation}) together with the trap parameters listed in Tab.~(\ref{tab:trap}), gives an ellipticity of $\epsilon_{\mathrm{trap}}=9.97(28)^\circ$. Such a large value is incompatible with the expected cylindrical symmetry of the trap. For comparison, Ref.~\cite{Berr2024_2} reported $\epsilon_{\mathrm{trap}}=0.7374(10)^\circ$ for the same trap as shown in Figs.~\ref{fig:setup} (b) and (c), when previously operated in a liquid-helium magnet with a 160~mm bore. In that work, the right-hand side of Eq.~(\ref{eq:misalignment_correlation}) was negative, and the approximation $\theta_{\mathrm{trap}}\ll\epsilon_{\mathrm{trap}}$ was adopted. Assuming the same ellipticity reported in Ref.~\cite{Berr2024_2} and using the measurements obtained at different trap depths (Tab.~(\ref{tab:trap})), one instead finds $\theta_{\mathrm{trap}}\approx1.49(8)^\circ$. Thus, the present analysis constrains the trap misalignment to the range $1.49(8)^\circ \lesssim \theta_{\mathrm{trap}} \lesssim 4.91(51)^\circ$. Even the lower bound indicates a substantial misalignment and therefore a dedicated alignment system is currently being developed to enable fine adjustment of the trap orientation with respect to the magnetic-field axis by moving the magnet.

\section{Linearization of power supply artifacts}
\label{app:newton_raphson}
\begin{figure}[b]
\centering
\includegraphics[width=0.50\textwidth]{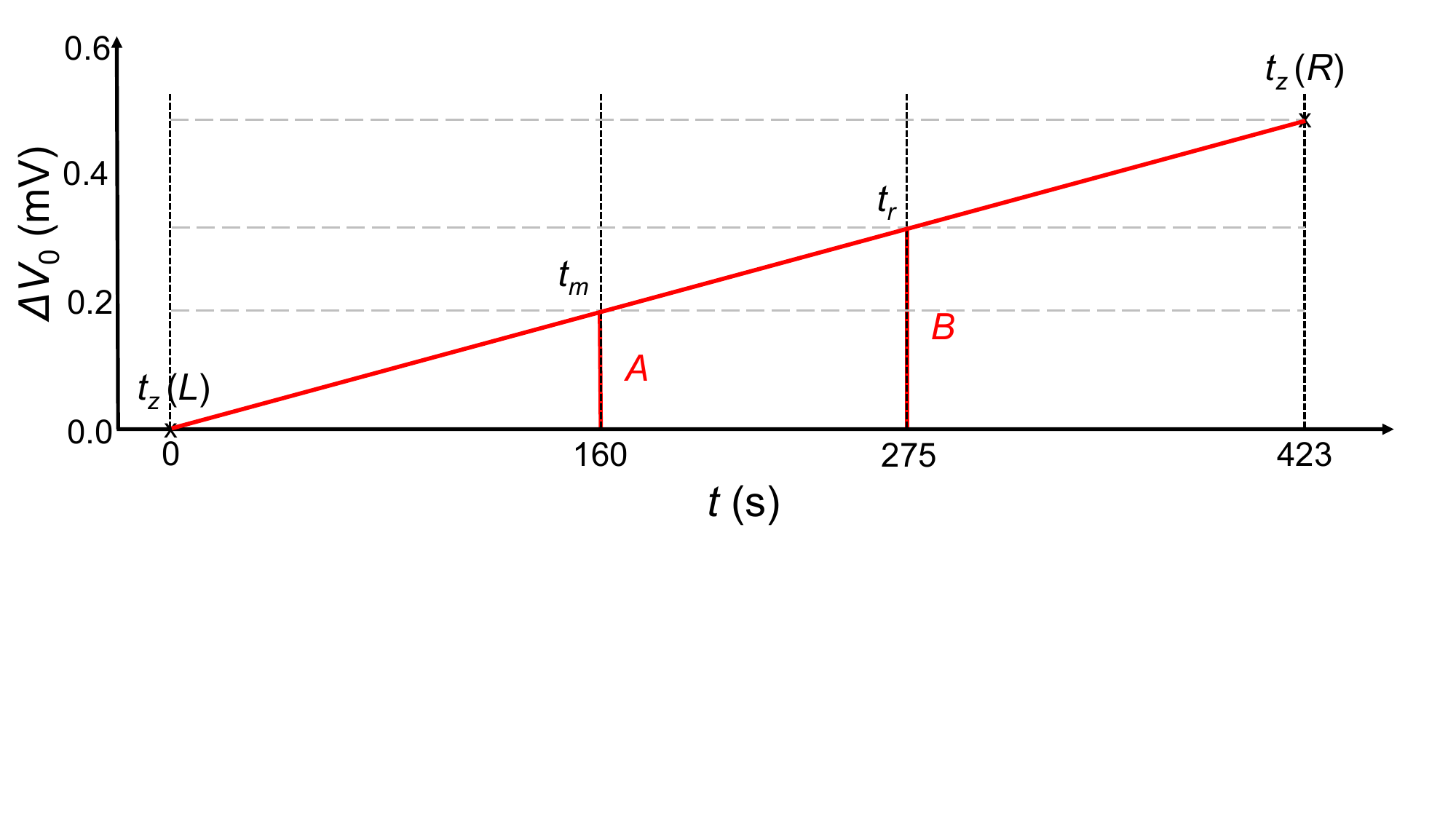}
\vspace{-25mm}
\caption{Schematic representation of the interpolation of $V_0$ between two neighboring axial measurements at $t_{z}(L)$ and $t_{z}(R)$ to obtain the potentials $V_0(t_m)$ and $V_0(t_r)$ used in the correction procedure. The vertical axis shows the deviation $\Delta V_0 = V_0 - 30\,\mathrm{V}$ in mV, illustrating the typical magnitude of the observed fluctuations. The horizontal axis indicates the midpoints of the three eigenfrequency measurements within a block.}
\label{fig:TDL_scheme}
\end{figure}

The variation of the trapping potential within a measurement block is assumed to be linear in time. Based on this assumption, a trap-depth linearization (TDL) is applied to estimate and correct artifacts arising from power-supply fluctuations while preserving genuine magnetic-field variations. The three eigenfrequencies of each triplet are referenced to the midpoints of their respective measurements, denoted by $t_z$, $t_m$, and $t_r$. The trapping potential is then obtained from the axial frequency via
\begin{equation}
V_0(t_z)
=
\frac{m d_0^2}{q}\,(2\pi)^2\nu_z^2(t_z).
\end{equation}
For each triplet, $V_0(t)$ is linearly interpolated from the neighboring axial measurements to the magnetron and reduced-cyclotron midpoints, yielding $V_0(t_m)$ and $V_0(t_r)$ (Fig.~\ref{fig:TDL_scheme}). The corresponding trap-depth variations within the triplet are quantified by
\begin{equation}
A \equiv V_0^{(m)} - V_0^{(z)},
\qquad
B \equiv V_0^{(r)} - V_0^{(z)}.
\label{eq:A_B}
\end{equation}
\begin{figure}[t]
\centering
\includegraphics[width=0.47\textwidth]{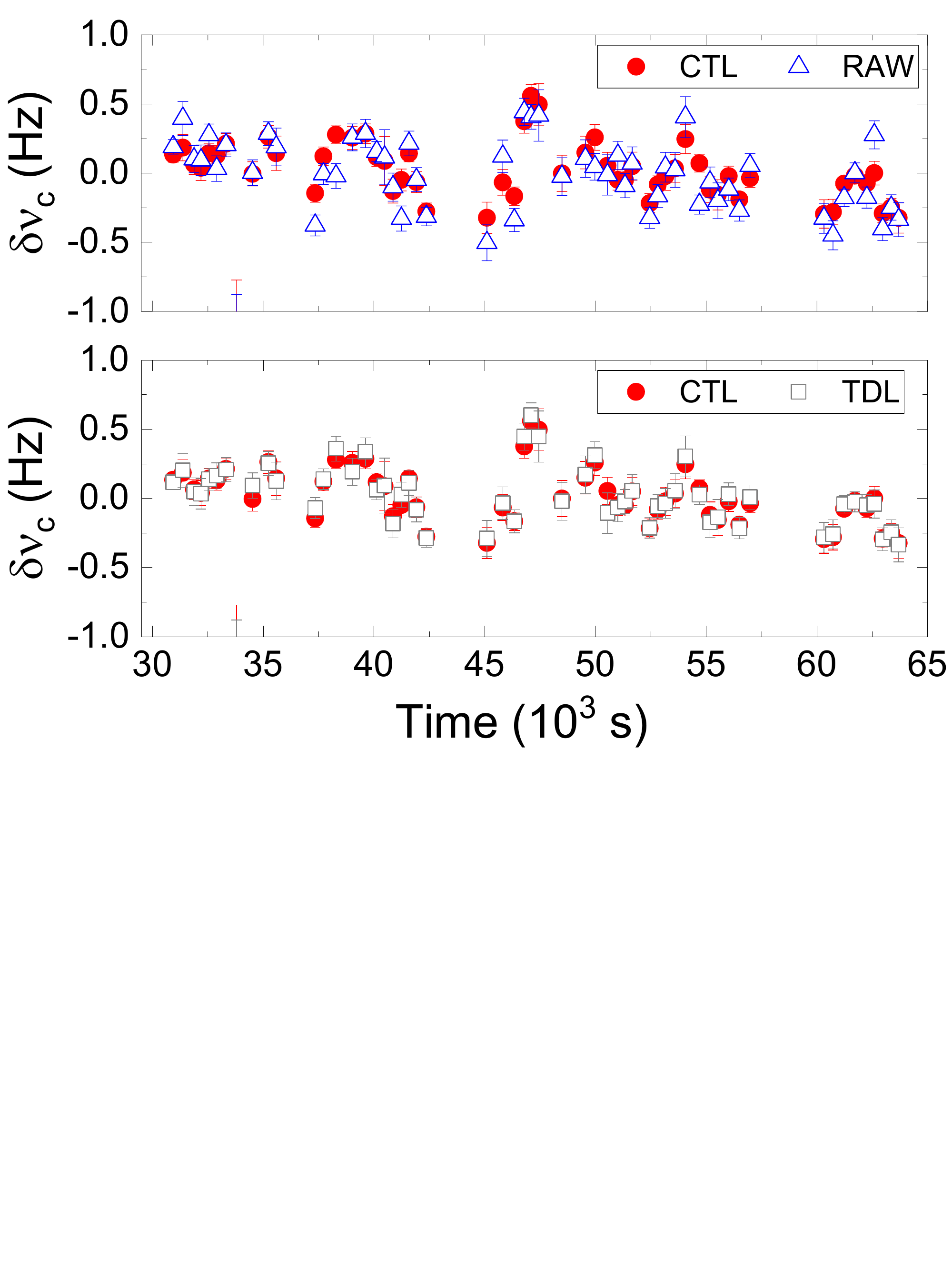}
\vspace{-47mm}
\caption{Upper panel: comparison between the cyclotron frequency obtained from the raw data and using the common-time linearization (CLT). Lower panel: comparison between CTL and TDL.}
\label{fig:TDL_comparision}
\end{figure}
In order to quantify the effect of trap-depth variations, we distinguish between the measured cyclotron frequency, $\nu_c^{\mathrm{meas}}$, obtained directly from the measured eigenfrequency triplet (raw data without linearization), and the corrected frequency, $\nu_c^{\mathrm{mod}}$, corresponding to considering a constant trapping potential throughout the measurement block. We assume
\begin{equation}
\nu_z^{\mathrm{meas}} = \nu_z^{\mathrm{mod}} = \nu_z,
\end{equation}
since the axial frequency depends only on the electric field and is measured first, thereby defining the electric-field reference for the corresponding block (see Fig.~\ref{fig:TDL_scheme}). The radial frequencies are then corrected for the interpolated trap-depth variations within the block to reconstruct $\nu_c^{\mathrm{model}}$. Assuming the linear temporal variation of the electric field (Fig.~\ref{fig:TDL_scheme}), we define 
\begin{equation} \Delta_A = \sqrt{ (\nu_c^{\mathrm{mod}})^2 - \frac{1}{(2\pi)^2}\frac{2q}{m d_0^2} \left[ V_0(t_z)+A \right] }, \label{eq:DeltaA} 
\end{equation} 
and 
\begin{equation} 
\Delta_B = \sqrt{(\nu_c^{\mathrm{mod}})^2 - \frac{1}{(2\pi)^2}\frac{2q}{m d_0^2} 
\left[ V_0(t_z)+B \right]}, \label{eq:DeltaB} 
\end{equation}
so that
\begin{equation} 
\nu_-^{\mathrm{meas}} = \frac{1}{2} \left( \nu_c^{\mathrm{mod}} - \Delta_A \right), \label{eq:omega_minus_meas} 
\end{equation} 
and 
\begin{equation} 
\nu_+^{\mathrm{meas}} = \frac{1}{2} \left( \nu_c^{\mathrm{mod}} + \Delta_B \right). \label{eq:omega_plus_meas} 
\end{equation}
The three frequencies are combined through the invariance theorem,
\begin{equation}
f(\nu_c^{\mathrm{mod}})=\sqrt{\nu_z^2(t_z)+\left(\nu_-^{\mathrm{mod}}\right)^2
+\left(\nu_+^{\mathrm{mod}}\right)^2},
\label{eq:f_omega}
\end{equation}
and the corrected cyclotron frequency is obtained from the nonlinear equation
\begin{equation}
f(\nu_c^{\mathrm{mod}})=\nu_c^{\mathrm{meas}},\label{eq:model}
\end{equation}
which is solved iteratively using a Newton--Raphson algorithm. It exploits the fact that both quantities are affected by the same systematic bias. The measured cyclotron frequency $\nu_c^{\mathrm{meas}}$ reflects the actual experimental conditions with a fluctuating power supply, which are consistently incorporated into $f(\nu_c^{\mathrm{mod}})$. The TDL approximation presents however two limitations. First, the power-supply fluctuations entering $\nu_c^{\mathrm{meas}}$ are not strictly linear in time, whereas the interpolation procedure necessarily assumes linear behavior. Second, $\nu_c^{\mathrm{meas}}$ contains an additional contribution arising from magnetic-field variations within a measurement block, which is not included in $f(\nu_c^{\mathrm{true}})$. These are expected to be much smaller than the corresponding electric-field fluctuations over the duration of a block, so that using Eq.~(\ref{eq:model}) remains well justified. Figure~\ref{fig:TDL_comparision} shows a comparison of the linearization, presented in the paper, with $\nu_c^{\mathrm{meas}}$ (upper panel), and between the linearization and the TDL approximation described in this appendix, for d6 in Fig.~\ref{fig:long_term_drift}. The linearized and TDL data are in good agreement with a mean difference of $0.0482(84)$~Hz.

\section{Temporal correlations in magnetic field fluctuations} \label{sec:temp_magnetic_field}
To assess whether the excess fluctuations observed in the lower panel of Fig.~\ref{fig:long_term_drift} are purely random or exhibit temporal correlations, the Durbin--Watson test was applied resulting in a value of 0.985. This indicates significant positive serial correlation, implying that neighboring residuals tend to assume similar values rather than fluctuating randomly around zero and suggesting the presence of correlated short-term variations. The pairwise second-order structure function was employed to characterize their timescale and amplitude quantitatively. For each day, all pairs of measurements were grouped according to their time separation $\tau$, and the mean squared difference was calculated as
\begin{equation}
D(\tau)=\frac{1}{2}\left\langle\left[x(t+\tau)-x(t)\right]^2\right\rangle=C(0)-C(\tau),
\label{eq:structure_function}
\end{equation}
\begin{figure}[t]
\includegraphics[scale=0.35]{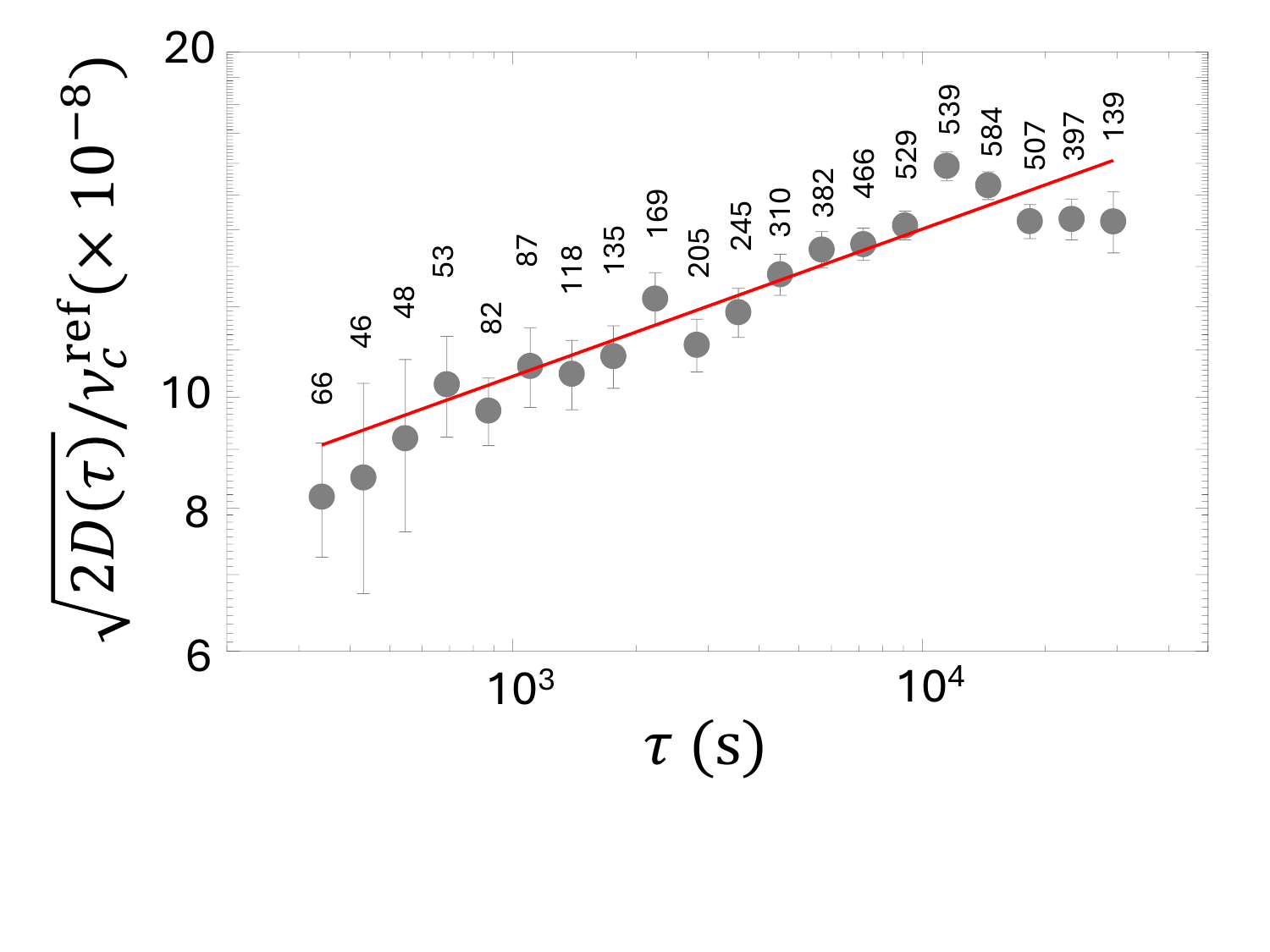}% Here is how to import EPS art
\vspace{-15mm}\caption{Second-order pairwise structure function combining data from all seven measurement days. Each day was analyzed individually, and the resulting structure functions were combined into a single representation. The $x$-axis denotes the time separation $\tau$ between two data points, while the $y$-axis, proportional to $\sqrt{2D(\tau)}/\nu_c^{\mathrm{ref}}$, represents the relative fluctuation amplitude occurring over that interval.  Numbers above the points indicate the number of pairs contributing to each value. A dependence, $D(\tau) \propto \tau^{\alpha}$ with $\alpha \approx 0.29$, is observed, indicating very weakly correlated fluctuations.\label{fig:structure_function}}
\end{figure}
where $x(t)=\nu_c(t)/\nu_c^{\mathrm{ref}}$ and $C(\tau)$ denotes the autocovariance. The resulting structure function is shown in Fig.~\ref{fig:structure_function}. Over time separations up to approximately 2.7~h, the structure function increases with increasing $\tau$ and can be described by a power-law dependence, $D(\tau)\propto\tau^\alpha$ with $\alpha \approx 0.29$. This behavior indicates weak but non-negligible temporal correlations: measurements close in time are more strongly correlated than those separated by longer intervals. The correlation gradually vanishes after approximately 2.7~h. The relatively small exponent indicates that the amplitude of the fluctuations grows only slowly with increasing time separation, consistent with weakly correlated magnetic-field variations possessing a limited temporal memory. To estimate the magnitude of these fluctuations, the analysis was restricted to ``reliable runs'', defined as sequences of at least five consecutive measurements for which the nearest neighbors were separated by less than 15~min. This criterion minimizes interpolation artifacts and reduces the influence of long sampling gaps. For each reliable run, the peak-to-peak variation was divided by its corresponding duration, $\Delta t_{\mathrm{pp}}$, to obtain a local fluctuation rate. The weighted mean of the absolute values yields
%\begin{equation}
%\left|\frac{\Delta \nu_c^{\mathrm{pp}}}{\Delta t_{\mathrm{pp}}}\right|
%=(1.43 \pm 0.10)\times10^{-2}\ \mathrm{Hz\,min^{-1}},
%\end{equation}
%or, expressed relative to the reference cyclotron frequency,
\begin{equation}
\frac{1}{\nu_{\mathrm{c}}} \left|\frac{\Delta \nu_c^{\mathrm{pp}}}{\Delta t_{\mathrm{pp}}}\right|
=5.32(37)\times10^{-9} \ \mathrm{min^{-1}}.
\end{equation}

It should be noted that the shortest timescale accessible with the present measurement scheme is approximately 5~min, corresponding to the time required to determine a complete eigenfrequency triplet. Although the function is different, the results are consistent with those shown in Fig.~\ref{fig:FFT_plot}.

%\bibliography{aipsamp}% Produces the bibliography via BibTeX.
\providecommand{\noopsort}[1]{}\providecommand{\singleletter}[1]{#1}%

\end{document}